\documentstyle[aps,pre,twocolumn,graphicx,psfig]{revtex}

\def\lessgtr{\raise2.5pt\hbox{$<$}\llap{\lower2.5pt\hbox{$>$}}}
\def\gtrless{\raise2.5pt\hbox{$>$}\llap{\lower2.5pt\hbox{$<$}}}

\newcommand{\be}{\begin{equation}}
\newcommand{\ee}{\end{equation}}
\newcommand{\bea}{\begin{eqnarray}}
\newcommand{\eea}{\end{eqnarray}}

\begin{document}

\draft

\title{The mean-squared displacement of a molecule
moving in a glassy system}
\author{S.-H. Chong, W.~G{\"o}tze, and M.~R. Mayr}
\address{Physik-Department, Technische Universit{\"a}t M{\"u}nchen,
85747 Garching, Germany}
\date{Phys. Rev. E, in print}
\maketitle

\begin{abstract}

The mean-squared displacement (MSD) of a hard sphere and of a dumbbell molecule
consisting of two fused hard spheres immersed in a dense
hard-sphere system is
calculated within the mode-coupling theory for ideal liquid-glass transitions.
It is proven that the velocity correlator, which is 
the second time derivative of the MSD, is the negative of a
completely monotone function for times within the structural-relaxation
regime. 
The MSD is found to exhibit a large time interval for structural relaxation
prior to the onset of the $\alpha$-process which cannot be described
by the asymptotic formulas for the mode-coupling-theory-bifurcation dynamics.
The $\alpha$-process for molecules with a large elongation is shown to
exhibit an anomalously wide cross-over interval between the end of the
von-Schweidler decay and the beginning of normal diffusion. 
The diffusivity of the molecule is predicted to vary non-monotonically
as function of its elongation.

\bigskip

\end{abstract}

\section{INTRODUCTION}

The mean-squared displacement $\delta r^{2}(t)$ (MSD) is a very transparent
concept for the discussion of liquid dynamics~\cite{Hansen86}.
For long times $t$, $\delta r^{2}(t)$ increases 
proportional to $t$ and to the diffusion constant of the fluid.
In an ideal solid, on the other hand, 
the long-time limit of $\delta r^{2}(t)$ is a finite
number characterizing the square of the particle's localization length.
Therefore, the long-time behavior of $\delta r^{2}(t)$ depends sensitively
on control parameters like density or temperature if the system is
close to a liquid-glass-transition point.
The MSD is thus particularly well suited to study glass-transition precursors.
It can be measured by incoherent inelastic-neutron-scattering
experiments.
However, $\delta r^{2}(t)$ has to be extracted as small-wave-number limit
of the intermediate scattering function~\cite{Hansen86}, and this makes it
very difficult to produce accurate data for large time intervals.
Dynamic light-scattering spectroscopy and sample-preparation techniques 
for colloidal suspensions made great progress in recent years.
It was demonstrated that very informative data for the MSD of hard-sphere
colloids near the glass transition can be obtained~\cite{Megen98};
and promising results for this system have also been measured by 
direct-imaging techniques~\cite{Kegel00,Weeks00}.
Molecular-dynamics simulations are well suited to get accurate data for the
MSD for liquids near the glass transition.
This was demonstrated for a binary
Lennard-Jones mixture~\cite{Kob95},
for a liquid of diatomic molecules~\cite{Kaemmerer98}, 
for models for the van-der-Waals liquid 
orthoterphenyl~\cite{Kudchadkar95,Mossa00}, 
for the SPC/E model for water~\cite{Gallo96},
for a hard-sphere-colloid model~\cite{Doliwa99},
and for a model of silica~\cite{Horbach99}.
In this paper, general features and some
quantitative results for the evolution of the glassy
dynamics as exhibited by the MSD will be considered within the
mode-coupling theory (MCT) for ideal liquid-glass transitions.

The basic version of MCT is built on approximately derived closed
equations of motion for the auto-correlation functions of density
fluctuations.
The essential input informations are the equilibrium structure
factors which are anticipated to vary smoothly with the system's
control parameters.
At certain critical values for the latter,
there occur bifurcations from solutions for an ergodic liquid to
ones for an amorphous non-ergodic solid.
Thus, MCT deals with a model for ideal liquid-glass transitions.
The transition implies a novel dynamical scenario.
Its features have been worked out by asymptotic solution of the
equations.
The leading-order asymptotic formulas establish universal
results for the MCT transition like scaling laws, power-law
divergencies for the time scales, and anomalous exponents for the
spectra~\cite{Goetze91b,Schilling94}.
Many tests of the relevance of the MCT results for the
explanation of the dynamics of glass-forming liquids have been
performed, which are reviewed to some extent in 
Ref.~\cite{Goetze99}.
Let us only mention here the recent analysis of data measured
for propylene carbonate~\cite{Wuttke00},
studies by means of the 
optical Kerr effect~\cite{Torre00,Hinze00}, 
and the analysis of simulation data 
for a binary Lennard-Jones liquid~\cite{Gleim00}
and for silica~\cite{Sciortino01}.
The outcome of these tests qualifies MCT as a candidate for a theory of
glassy dynamics and it seems justified to continue the
preceding studies by unfolding some of its implications for the MSD.

The intention of this paper is to identify further MCT results for 
future tests of this theory.
The previous work on the hard-sphere system 
(HSS)~\cite{Fuchs92b,Fuchs95,Franosch97,Fuchs98,Goetze00}
shall be continued by analyzing in detail the MSD for a tagged particle.
The work on the MCT for molecular systems~\cite{Chong01} will be
extended by evaluating the MSD for the interaction sites of a symmetric
dumbbell consisting of two fused hard spheres as well as for
the molecule's center.
The paper is structured as follows.
In Sec.~\ref{sec:2}, the equations to be solved are listed and the
concepts to be used to discuss the results are described.
Section~\ref{sec:3} presents the results for the MSD and the analysis
of its properties.
In Sec.~\ref{sec:4}, the findings are summarized.

\section{BASIC FORMULAS}
\label{sec:2}

\subsection{Description of the system}
\label{subsec:description}

In this section, the systems to be studied and the functions to be used
for a description of their dynamics shall be defined.
A system of $N$ atoms of mass $m$
distributed with density $\rho$ is considered as solvent. 
The points in configuration space are specified by the 
particle positions ${\vec r}_\kappa$, $\kappa = 1,2, \cdots, N$. 
The basic variables for the description of the
structure are the density fluctuations with wave vector ${\vec q}$,
$\rho_{\vec q} = \sum_\kappa \exp (i {\vec q} \cdot {\vec r}_\kappa)$. If
$\langle \cdots \rangle$ denotes canonical averaging for
temperature $T$, the structure factor is $S_q = \langle \mid
\rho_{\vec q} \mid^2 \rangle / N$, where $q = \mid \vec q \mid$ is
the wave number. 
As simplest example for a solute, a tagged atom of mass $m_S$ and 
position ${\vec r}_S$ shall be considered. 
The distribution of the atom is described by the density
fluctuation $\rho_{\vec q}^S = \exp (i {\vec q} \cdot {\vec r}_S)$. 
The solute-solvent interaction shall be characterized by the direct
correlation function 
$c_q^S = \langle \rho_{\vec q}^{S *} \rho_{\vec q} \rangle / (\rho S_q)$.
As a more complicated solute, a symmetric rigid diatomic molecule
shall be chosen. 
Its position is specified by two interaction sites 
$(\vec r_A, \vec r_B)$, which have the same mass $m_{A}$. 
The position of the molecule can also be described by its
center ${\vec r}_C = ({\vec r}_A + {\vec r}_B)/2$ and the unit
vector $\vec e = ({\vec r}_A - {\vec r}_B) / L$, where
$L = \, \mid {\vec r}_A - {\vec r}_B \mid$. 
The configuration variables for the molecule can be built with the two
density fluctuations 
$\rho_{\vec q}^a = \exp (i {\vec q} \cdot {\vec r}_a)$;
$a = A$, $B$. 
Equivalently, one can use the number fluctuations
$\rho^{N}_{\vec q} = (\rho_{\vec q}^A + \rho_{\vec q}^B) / \sqrt{2}$
and the ``charge'' fluctuations 
$\rho^{Z}_{\vec q} = (\rho_{\vec q}^A - \rho_{\vec q}^B) / \sqrt{2}$. 
The solute-solvent interaction can be characterized by the
direct interaction-site-solvent correlations 
$c_q^A = c_q^B = \langle \rho_{\vec q}^{A *} \rho_{\vec q} \rangle / 
[\rho S_{q} w_{q}^{N}]$~\cite{Chandler72}.
Here $w_{q}^{N,Z} = 1 \pm \sin(qL)/qL$ denote the intra-molecular
structure factors.

Three kinds of mean-squared displacement functions of time $t$ shall
be discussed, 
$\delta r_x^2 (t) = \langle [{\vec r}_x (t) - {\vec r}_x (0)]^2 \rangle$.
Here and in the following, the label $x=S$, $C$ and $A$ refers to the
position of a tagged particle, of the center of the molecule, and of the
atomic center in the molecule, respectively.
It will be more convenient to use the following abbreviation
\begin{equation}
\Delta_{x}(t) = {\textstyle \frac{1}{6}} \, 
\delta r_{x}^{2}(t), \quad
x=S, C, A.
\label{eq:Delta-x-def}
\end{equation}
The MSD of the molecule's constituents can be decomposed into one
contribution due to translation of the center and one due to
reorientation of the axis~\cite{Chong01}
\begin{equation}
\Delta_{A}(t) = \Delta_{C}(t) + {\textstyle \frac{1}{12}} \,
L^{2} \, [1 - C_{1}(t)],
\label{eq:Delta-A-decom}
\end{equation}
where
\begin{equation}
C_{1}(t) = \langle {\vec e}(t) \cdot {\vec e} \, \rangle
\label{eq:C1-def}
\end{equation}
is the dipole correlator.
The time derivatives of the MSD provide the velocity-auto-correlation
function~\cite{Hansen86}.
Let us consider the one for the velocity
${\vec v}_{S}(t)$ of the tagged particle only,
$K_{S}(t) = \langle {\vec v}_{S}(t) \cdot {\vec v}_{S} \rangle$,
where
\begin{equation}
\partial_{t}^{2} \Delta_{S}(t) = {\textstyle \frac{1}{3}} \, K_{S}(t).
\label{eq:Delta-S-prop}
\end{equation}

The non-trivial time dependence of ${\vec r}_{x}(t) - {\vec r}_{x}(0)$
comes about since the forces on the solute fluctuate in time;
and this is caused by the density fluctuations of the solvent and by
the fluctuations of the probability density of the solute constituents.
These quantities are described by the density correlator
$\phi_{q}(t) = \langle \rho_{\vec q}(t)^{*} \rho_{\vec q} \rangle /
\langle \mid \rho_{\vec q} \mid^{2} \rangle$
of the solvent, by the tagged-particle-density correlator
$\phi^{S}_{q}(t) = \langle \rho^{S}_{\vec q}(t)^{*} \rho^{S}_{\vec q} \rangle$,
and by the molecule's correlators
$\phi^{N,Z}_{q}(t) = \langle \rho^{N,Z}_{\vec q}(t)^{*} \rho^{N,Z}_{\vec q} \rangle /
w_{q}^{N,Z}$. 
These correlators also determine the desired functions,
since~\cite{Hansen86,Chong01}:
\begin{mathletters}
\label{eq:phi-x-qnull}
\bea
\phi^{S}_{q}(t) &=&
1 - q^{2} \, \Delta_{S}(t) + O(q^{4}),
\label{eq:phi-x-qnull-a}
\\
\phi^{N}_{q}(t) &=&
1 - q^{2} \, \Delta_{C}(t) + O(q^{4}),
\label{eq:phi-x-qnull-b}
\\
\phi^{Z}_{q}(t) &=&
C_{1}(t) + O(q^{2}).
\label{eq:phi-x-qnull-c}
\eea
\end{mathletters}

\subsection{MCT approximations}

In this section, those equations 
shall be listed which have to be solved numerically.
Within the Zwanzig-Mori theory, an exact equation of motion can be
formulated for the density correlator:
$\partial_{t}^{2} \phi_{q}(t) + \Omega_{q}^{2} \, \phi_{q}(t) + 
\int_{0}^{t} dt^{\prime} \, M_{q}(t-t^{\prime}) \, 
\partial_{t^{\prime}} \phi_{q}(t^{\prime}) = 0$.
Here $\Omega_{q}^{2} = v^{2} q^{2} / S_{q}$ 
with $v$ denoting the thermal velocity specifies
a characteristic frequency $\Omega_{q}$, and $M_{q}(t)$ denotes
a fluctuating-force correlator called the relaxation 
kernel~\cite{Hansen86}.
Within MCT, the kernel is split into a regular part 
$M_{q}^{\rm reg}(t)$
dealing with normal-liquid effects and a mode-coupling kernel
$\Omega_{q}^{2} m_{q}(t)$ describing the cage effect.
If one introduces an operator ${\cal R}$ for the regular dynamics by
${\cal R} \phi_{q}(t) = 
[ \partial_{t}^{2} \phi_{q}(t) + 
\int_{0}^{t} dt^{\prime} \, M^{\rm reg}_{q}(t-t^{\prime}) \, 
\partial_{t^{\prime}} \phi_{q}(t^{\prime}) ] / \Omega_{q}^{2}$,
one can write
\begin{equation}
{\cal R} \phi_{q}(t) + \phi_{q}(t) +
\int_{0}^{t} dt^{\prime} \, m_{q}(t-t^{\prime}) \, 
\partial_{t^{\prime}} \phi_{q}(t^{\prime}) = 0.
\label{eq:GLE-v}
\end{equation}
The crucial step in the derivation is the application of Kawasaki's
factorization approximations to express the kernel 
$m_{q}(t)$ as mode-coupling functional ${\cal F}_{q}$ of the
correlators
\begin{mathletters}
\label{eq:MCT-v}
\bea
m_{q}(t) &=& {\cal F}_{q}[ \phi_{k}(t)],
\label{eq:MCT-v-a}
\\
{\cal F}_{q}[\tilde{f}_{k}] &=& \frac{1}{2 (2\pi)^{3}}
\int d{\vec k} \, V({\vec q}; {\vec k}, {\vec p} \,) \,
\tilde{f}_{k} \, \tilde{f}_{p}.
\label{eq:MCT-v-b}
\eea
Here $\vec p$ abbreviates $\vec q - \vec k$. 
The coefficients $V ({\vec q}; {\vec k}, {\vec p} \,)$ 
are given in terms of the structure factor~\cite{Bengtzelius84}.
\end{mathletters}

None of the MCT results for structural relaxations, in particular
none of the universal results to be cited in the following 
Sec.~\ref{subsec:universal},
depend on the model for $M_{q}^{\rm reg}(t)$.
This holds up to the choice of some time scale to be denoted below
as $t_{0}$.
But, the kernel $M_{q}^{\rm reg}(t)$ shall be specified in order to have
controllable quantitative results for all times.
Specifically, a model with $M_{q}^{\rm reg}(t) \equiv 0$
shall be chosen.
The operator ${\cal R}$ shall be complemented by an index 
${\rm H}$ indicating that a Hamiltonian dynamics is considered for the
short-time motion
\begin{mathletters}
\label{eq:R-model}
\begin{equation}
{\cal R}^{\rm H} \phi_{q}(t) = \partial_{t}^{2} \phi_{q}(t) \, / \, \Omega_{q}^{2}.
\label{eq:R-model-a}
\end{equation}
This model overemphasizes oscillation features.
A more realistic model would include at least some 
friction term as it is caused for low frequency phenomena by binary
collision events.
But, no detailed proposals for a treatment of such effects have
been made so far within MCT. 
Some results also will be presented for a simplified colloid model.
Here, the inertia term from Eq.~(\ref{eq:R-model-a}) is neglected
and the regular term is chosen as a $q$-independent white noise
kernel.
It is explained in more detail in Appendix~\ref{appen:A}
that this model corresponds to the conventional treatment of colloids
by coarse-graining the time over intervals of the duration of 
collisions of the solvent molecules with the mesoscopic colloid particles.
As a result, the short-time motion is treated by a Brownian dynamics
\begin{equation}
{\cal R}^{\rm B} \phi_{q}(t) = \tau_{q} \, \partial_{t} \phi_{q}(t).
\label{eq:R-model-b}
\end{equation}
Here $\tau_{q} = S_{q} / (D_{0} q^{2})$ with $D_{0}$ denoting the
single-particle-diffusion constant.

Equations~(\ref{eq:GLE-v}) and (\ref{eq:R-model}) hold analogously for
the solute correlators $\phi^{x}_{q}(t)$, $x=S$, $N$ and $Z$.
\end{mathletters}
One gets $\Omega_{q}^{S \, 2} = v_{S}^{2} q^{2}$ with $v_{S}$ denoting
the tagged particle thermal velocity.
The relaxation time for the Brownian motion is
$\tau^{S}_{q} = 1 / (D^{S}_{0} q^{2})$ with $D^{S}_{0}$ denoting
the tagged particle short-time diffusivity.
The more involved expressions for the characteristic frequencies
$\Omega^{N,Z}_{q}$ can be found in Ref.~\cite{Chong01}.
Brownian dynamics shall not be considered for the dumbbell molecule.
The fluctuating-force kernels are
functionals of the correlators $\phi^{x}_{q}(t)$ and 
$\phi_{q}(t)$:
\begin{mathletters}
\label{eq:MCT-x}
\bea
m_{q}^{x}(t) &=& {\cal F}_{q}^{x}[ \phi_{k}^{x}(t), \phi_{p}(t)],
\label{eq:MCT-x-a}
\\
{\cal F}_{q}^{x}[ \tilde{f}_{k}^{x}, \tilde{f}_{p}] &=&
\frac{1}{(2 \pi)^{3}}
\int d{\vec k} \, V^{x}({\vec q}; {\vec k}, {\vec p} \,) \,
\tilde{f}_{k}^{x} \, \tilde{f}_{p},
\label{eq:MCT-x-b}
\eea
for $x=S, N, Z$. 
Again, ${\vec p}$ abbreviates ${\vec q}-{\vec k}$ and 
the coefficients $V^{x}({\vec q}; {\vec k}, {\vec p} \,)$ are
given by $S_{q}$ and the direct correlation 
functions~\cite{Chong01,Bengtzelius84}.
\end{mathletters}
It is cumbersome to calculate the required $q \to 0$ limits
in Eqs.~(\ref{eq:phi-x-qnull})
numerically from numerical solutions for $\phi^{x}_{q}(t)$.
It is more adequate to carry out the limit analytically in the equations
of motion for $\phi^{x}_{q}(t)$ so that one gets equations of motion
for the desired functions.
The non-trivial parts of these equations are convolution integrals 
defined with the $q \to 0$ limits of the kernels
$m^{x}_{q}(t)$.
One gets for the dipole correlator
\bea
& &
\partial_{t}^{2} C_{1}(t) + 2 v_{\rm R}^{2} \, C_{1}(t) 
\nonumber \\
& & \quad \quad \quad
+ \,
2 v_{\rm R}^{2} \, \int_{0}^{t} dt^{\prime} \,
m_{Z}(t - t^{\prime}) \, \partial_{t^{\prime}} C_{1}(t^{\prime}) = 0,
\label{eq:GLE-C1}
\eea
where $v_{\rm R}$ is the thermal angular velocity of the 
molecule~\cite{Chong01}.
From the equation for $\phi^{S}_{q}(t)$ one gets a Zwanzig-Mori
equation for the velocity correlator~\cite{Goetze91b}
\begin{equation}
\partial_{t} K_{S}(t) + v_{S}^{2} \, 
\int_{0}^{t} dt^{\prime} \,
m_{S}(t-t^{\prime}) \, K_{S}(t^{\prime}) = 0.
\label{eq:GLE-K}
\end{equation}
Integrating twice over $t$ one gets with the aid of 
Eq.~(\ref{eq:Delta-S-prop}) and the initial conditions
$\Delta_{S}(0) = 0$ and $\partial_{t} \Delta_{S}(0) = 0$,
\begin{mathletters}
\label{eq:GLE-Delta-S}
\begin{equation}
\mbox{H:} \quad
\partial_{t} \Delta_{S}(t) - v_{S}^{2} \, t + v_{S}^{2} \,
\int_{0}^{t} dt^{\prime} \, m_{S}(t-t^{\prime}) \, \Delta_{S}(t^{\prime}) = 0.
\label{eq:GLE-Delta-S-a}
\end{equation}
The corresponding equation for Brownian short-time 
dynamics~\cite{Fuchs98} is derived in Appendix~\ref{appen:A}:
\begin{equation}
\mbox{B:} \quad
\Delta_{S}(t) - D_{0}^{S} \, t +
D_{0}^{S} \, 
\int_{0}^{t} dt^{\prime} \, m_{S}(t-t^{\prime}) \, \Delta_{S}(t^{\prime}) = 0.
\label{eq:GLE-Delta-S-b}
\end{equation}
The same procedure leads to the equation of motion
for the MSD of the center
\end{mathletters}
\begin{equation}
\mbox{H:} \quad
\partial_{t} \Delta_{C}(t) - v_{\rm T}^{2} \, t + v_{\rm T}^{2} \,
\int_{0}^{t} dt^{\prime} \, m_{N}(t-t^{\prime}) \, \Delta_{C}(t^{\prime}) = 0,
\label{eq:GLE-Delta-C}
\end{equation}
where the kernel is denoted by $m_N (t)$ and $v_{\rm T}$ is the thermal
velocity for the molecule's translation~\cite{Chong01}.
The problem of the molecule's dynamics reduces to that of a
tagged atom with $m_{S} = 2 m_{A}$, 
if the limit $L \to 0$ is considered. 

The kernels in the preceding Eqs.~(\ref{eq:GLE-C1})--(\ref{eq:GLE-Delta-C})
are given by mode-coupling functionals
\begin{mathletters}
\label{eq:MCT-x-qnull}
\bea
m_{x}(t) &=& {\cal F}^{x}[\phi_{k}^{x}(t), \phi_{p}(t)],
\label{eq:MCT-x-qnull-a}
\\
{\cal F}^{x}[\tilde{f}_{k}^{x}, \tilde{f}_{p}] &=&
\frac{1}{6 \pi^{2}}
\int_{0}^{\infty} dk \, k^{4} \, \rho S_{k} \, v^{x}(k) \,
\tilde{f}_{k}^{x} \, \tilde{f}_{k},
\label{eq:MCT-x-qnull-b}
\eea
for $x=S, N, Z$.
Here $v^{S}(k) = c_{k}^{S \, 2}$~\cite{Bengtzelius84},
$v^{N}(k) = 2 c_{k}^{A \, 2} w_{k}^{N}$ and
$v^{Z}(k) = (L^{2} / 6) c_{k}^{A \, 2} w_{k}^{Z}$~\cite{Chong01}.
\end{mathletters}
A fluctuating force with vanishing wave vector can couple to
density fluctuations of the solvent for all wave vectors $\vec k$
provided the atom or molecule can absorb the recoil with wave
vector $- \vec k$. 
Therefore, one needs the superposition of
density correlators $\phi_k (t)$ and $\phi_k^x (t)$ for all wave
numbers $k$ for the calculation of the kernels $m_x (t)$.

\subsection{Universal results}
\label{subsec:universal}

Universal properties of the MCT-glass-transition scenario are
formulated by the leading-order asymptotic expressions
for the long-time dynamics for states near the transition points.
This paper focuses on features beyond the 
universal ones, but
the universal formulas shall be used as reference.
In this section, those formulas~\cite{Goetze91b} shall be compiled 
which are needed in the following Sec.~\ref{sec:3}
for the description of the results.  

The equilibrium structure of the system may depend on, say $n$,
control parameters which can be combined to a control-parameter
vector $V$.
A separation parameter $\sigma(V)$, 
a smooth function of $V$, can be defined with the aid of the
mode-coupling functional ${\cal F}_{q}$.
For states with control parameters $V$ such that $\sigma < 0$,
correlation functions decay to zero: $\phi_{q}(t \to \infty)=0$.
But for states with $\sigma > 0$, density fluctuations exhibit
spontaneous arrest: $\phi_{q}(t \to \infty) = f_{q} > 0$.
The Debye-Waller factor $f_{q}$ is to be evaluated from the
mode-coupling functional ${\cal F}_{q}$ in Eqs.~(\ref{eq:MCT-v})
via the equation 
$f_{q}/(1-f_{q}) = {\cal F}_{q}[f_{k}]$~\cite{Bengtzelius84}.
The set of critical points $V^{c}$,
defined by $\sigma(V^{c})=0$, separates liquid states from
glass states.
This result holds for all correlators 
$\phi_{A}(t) = \langle A(t)^{*} A \rangle / \langle | A |^{2} \rangle$
of variables $A$ coupling to density fluctuations.
While $\phi_{A}(t \to \infty)$ vanishes for the liquid,
generically, the limit $f_{A} = \phi_{A}(t \to \infty)$ is positive
for the glass.
If $A$ refers to $\rho_{\vec q}$, $\rho^{S}_{\vec q}$ or ${\vec e}$,
$f_{A}$ denotes the Debye-Waller factor $f_{q}$, 
the Lamb-M\"ossbauer factor $f_{q}^{S}$ or the Edwards-Anderson
parameter $f_{1}$, respectively, of the glass.
Crossing the transition points, the long-time limit
changes discontinuously from zero to the critical value $f_{A}^{c}$.
For the variation of $f_{A}$ upon approaching the transition from the
glass side, one gets for small $\sigma$ in leading order
\be
f_{A} = f_{A}^{c} + h_{A} \, \sqrt{ |\sigma| } \, / \, \sqrt{1 - \lambda}.
\label{eq:f-A}
\ee
Here $h_{A}$ is called the critical amplitude for variable $A$.
Every point $V^{c}$ is characterized by a number $\lambda$,
$0 < \lambda < 1$, which is called the exponent parameter.
The significance of $\lambda$ is explained below in connection with
Eqs.~(\ref{eq:phi-A-critical})--(\ref{eq:phi-A-von}).
It is a matter of convention not to incorporate 
$\sqrt{1 - \lambda}$ in the amplitude $h_{A}$,
in order to simplify some of the following formulas.
The quantities $f_{A}^{c}$, $h_{A}$ and $\lambda$ are calculated
from ${\cal F}_{q}$.
They are equilibrium quantities which are the same for Hamiltonian
and Brownian dynamics.

The parameter $\lambda$ determines an anomalous exponent $a$,
$0 < a < 1/2$, which is called the critical exponent.
There holds the equation 
$\Gamma(1-a)^{2}/\Gamma(1-2a) = \lambda$, where
$\Gamma$ denotes the gamma function.
The long-time decay at the critical point is given, up to corrections
of order $t^{-2a}$, by the power law
\be
\phi_{A}(t) = f_{A}^{c} + h_{A} \, (t_{0}/t)^{a}, \quad
\sigma = 0, \quad
t/t_{0} \gg 1.
\label{eq:phi-A-critical}
\ee
The $A$-independent time $t_{0}$ is the relevant microscopic scale for the
bifurcation dynamics.
It depends on all details of the transient dynamics as well as
on the mode-coupling functionals for parameters at the transition point.

The first scaling law of MCT deals with the dynamics for small $\sigma$
in a time interval where $\eta = \phi_{A}(t) - f_{A}^{c}$ is small.
In a leading expansion for $\sigma \to 0$ and $\eta \to 0$, one gets
\begin{mathletters}
\label{eq:phi-A-first}
\bea
\phi_{A}(t) &=& 
f_{A} + h_{A} \, G(t),
\label{eq:phi-A-first-a}
\\
G(t) &=&
\sqrt{ | \sigma | } \, g_{\pm}(t/t_{\sigma}), \quad
\sigma \gtrless 0,
\label{eq:phi-A-first-b}
\\
t_{\sigma} &=&
t_{0} \, / \, | \sigma |^{\delta}, \quad
\delta = 1 / 2a.
\label{eq:phi-A-first-c}
\eea
The functions $g_{\pm}(\hat{t})$ are determined by $\lambda$.
\end{mathletters}
Thus, the control-parameter dependence of the dynamics is
determined by the correlation scale  $\sqrt{|\sigma|}$
and by the first critical time scale $t_{\sigma}$.
One gets $g_{\pm}(\hat{t} \to 0) = 1 / \hat{t}^{a}$, so that
Eq.~(\ref{eq:phi-A-critical}) is reproduced for fixed large $t$
if $\sigma$ tends to zero.
Since $g_{+}(\hat{t} \to \infty) = 1 / \sqrt{1-\lambda}$, 
also Eq.~(\ref{eq:f-A}) is reproduced. 

There holds $g_{-}(\hat{t} \to \infty) = - B \hat{t}^{b} + O(1/\hat{t}^{b})$.
The anomalous exponent $b$, $0 < b \le 1$, which is called 
the von Schweidler exponent, is to be calculated from the equation
$\Gamma(1+b)^{2} / \Gamma(1+2b) = \lambda$.
The constant $B$ is of order unity.
Substituting this result into Eqs.~(\ref{eq:phi-A-first}) one gets 
von Schweidler's law for the decay of the liquid correlator below the
plateau $f_{A}^{c}$
\begin{mathletters}
\label{eq:phi-A-von}
\be
\phi_{A}(t) = f_{A}^{c} - h_{A} \, (t/t_{\sigma}^{\prime})^{b}, \quad
t_{\sigma} \ll t, \quad
\sigma \to -0.
\label{eq:phi-A-von-a}
\ee
The control-parameter dependence is described
by the second critical time scale
\be
t_{\sigma}^{\prime} = t_{0} \, B^{-1/b} \, / \, |\sigma|^{\gamma}, \quad
\gamma = (1/2a) + (1/2b).
\label{eq:phi-A-von-b}
\ee

Following the terminology of the glass-transition literature, the decay
of $\phi_{A}(t)$ below the plateau $f_{A}^{c}$ is called the 
$\alpha$-process.
\end{mathletters}
For this process, there holds the second scaling law of MCT 
in leading order for $\sigma \to -0$:
\be
\phi_{A}(t) = \tilde{\phi}_{A}(\tilde{t}), \quad
\tilde{t} = t / t_{\sigma}^{\prime}, \quad
t_{\sigma} \ll t.
\label{eq:phi-A-second}
\ee
The control-parameter-independent shape function 
$\tilde{\phi}_{A}(\tilde{t})$ is to be evaluated from the mode-coupling
functionals at the critical points $V^{c}$. 
The differences of the dynamics as they are caused by different models
for the short-time dynamics merely enter via differences in the
scale $t_{0}$.
For short rescaled times $\tilde{t}$,
one gets 
$\tilde{\phi}_{A}(\tilde{t}) = f_{A}^{c} - h_{A} \tilde{t}^{b} + 
h_{A}^{\prime} \tilde{t}^{2b} + \cdots$,
so that Eq.~(\ref{eq:phi-A-von-a}) is reproduced.
The ranges of applicability of the first and the second scaling laws
overlap;
both scaling laws imply von Schweidler's law for 
$t_{\sigma} \ll t \ll t_{\sigma}^{\prime}$.

Suppose, the system is driven through the transition point $V^{c}$
by smooth variation of some parameters $\theta$ like the
temperature, the density or, for a colloid, the salt concentration 
of the solvent.
Let $\theta^{c}$ denote the value where $V(\theta^{c}) = V^{c}$.
Then one can write for $\sigma$, in leading order
for small $(\theta - \theta^{c})$, the expression
$\sigma = C_{\theta} \, (\theta - \theta^{c}) / \theta^{c}$.
The constant $C_{\theta}$ depends on the choice of $\theta$
and connects the distance parameter 
$\epsilon = (\theta-\theta^{c})/\theta^{c}$ with the relevant
separation parameter $\sigma$.

\subsection{The model}

Hard spheres of diameter $d$ shall be used as model for the solvent
atoms. For this case, all equilibrium quantities are specified by
the packing fraction $\varphi = \pi \rho d^3 /6$. 
The MCT model
for the hard-sphere system (HSS) shall be 
defined by two further technical assumptions.
First, the structure factor $S_q$ and the direct correlation
function $c_q$ are evaluated within the Percus-Yevick 
theory~\cite{Hansen86}. 
Second, the wave numbers are discretized to 100
equally spaced values $q d = 0.2, 0.6, 1.0, \cdots, 39.8$. The
details of the transformation of the functional in 
Eq.~(\ref{eq:MCT-v-b}) to a
polynomial in the 100 variables $\phi_{q}(t)$ 
can be found in Ref.~\cite{Franosch97}. 
Representative solutions are
shown in Refs.~\cite{Goetze00} and \cite{Franosch98}
for the Hamiltonian dynamics and in Ref.~\cite{Franosch97}
for the Brownian dynamics.
There is a liquid-glass transition at the critical
packing fraction 
$\varphi_c \approx 0.516$~\cite{Franosch97,Bengtzelius84}. 
For the exponent parameter one gets $\lambda = 0.735$, and this implies
\begin{mathletters}
\label{eq:MCT-parameters}
\be
a = 0.312, b = 0.583, B = 0.836,
\delta = 1.60, \gamma = 2.46.
\label{eq:MCT-parameters-a}
\ee
For the separation parameter $\sigma$ one gets in leading order
\be
\sigma = 1.54 \, \epsilon, \quad
\epsilon = (\varphi - \varphi_{c}) / \varphi_{c}.
\label{eq:MCT-parameters-b}
\ee
The microscopic time scales
$t_{0}$ for Hamiltonian dynamics~\cite{Franosch98} and Brownian 
dynamics~\cite{Franosch97}, respectively, are
\end{mathletters}
\be
t_{0}^{\rm H} = 0.0236 \, (d/v), \quad
t_{0}^{\rm B} = 0.00265 \, d^{2} / D_{0}.
\label{eq:t0}
\ee

As atomic solute, a tagged particle of the solvent shall be
considered, i.e., $m_S = m$ and $c_q^S$ is identical with the direct
correlation function $c_{q}$ of the HSS. 
As molecular solute,
a symmetric dumbbell of two fused hard spheres of diameter $d$ and mass 
$m_A = m$ shall be chosen. 
The elongation $\zeta = L/d$ shall be used
as control parameter of the solute. 
The wave numbers are chosen discrete as above. 
The direct correlation
function $c_q^A$ is expressed as series of contributions
$c_{\ell}(q), \, \ell = 0,1, \cdots$, obtained by expanding the
molecule-solvent correlations in spherical 
harmonics~\cite{Chong01}. The sum over $\ell$ is truncated at 
$\ell_{\rm co} = 8$. The $c_\ell(q)$ are evaluated with the Percus-Yevick 
theory~\cite{Franosch97b}. 
Representative results for 
molecule's correlators $\phi_q^N (t)$, $\phi_q^Z (t)$ and $C_1 (t)$ are
shown in Ref.~\cite{Chong01}, and 
for $\phi^{S}_{q}(t)$ for the Brownian dynamics in 
Ref.~\cite{Fuchs98}. 

The figures to be discussed below and the numbers to be mentioned
are evaluated for the above-specified model as follows.
First, for a representative set of packing fractions, 
Eqs.~(\ref{eq:GLE-v}) and (\ref{eq:MCT-v}) are solved for the
density correlators $\phi_{q}(t)$, and this for Hamiltonian
as well as for Brownian dynamics.
These correlators are used to define the kernels in 
Eqs.~(\ref{eq:MCT-x}), so that, as second step, the tagged
particle correlators $\phi^{S}_{q}(t)$ could be evaluated,
also for both examples for the short-time dynamics.
Furthermore, for every $\varphi$, 
the equations for molecule's correlators 
$\phi^{N,Z}_{q}(t)$ are solved for 10 values for the
elongation $\zeta$.
These results are substituted into 
Eqs.~(\ref{eq:MCT-x-qnull}) for the kernels $m_{x}(t)$
so that, as last step, 
Eqs.~(\ref{eq:GLE-C1})--(\ref{eq:GLE-Delta-C})
for the desired functions $\Delta_{x}(t)$, $C_{1}(t)$ and
$K_{S}(t)$ can be solved.

\section{RESULTS}
\label{sec:3}

\subsection{The diffusion-localization transition}

\begin{figure}
\centerline{\scalebox{0.80}{\includegraphics{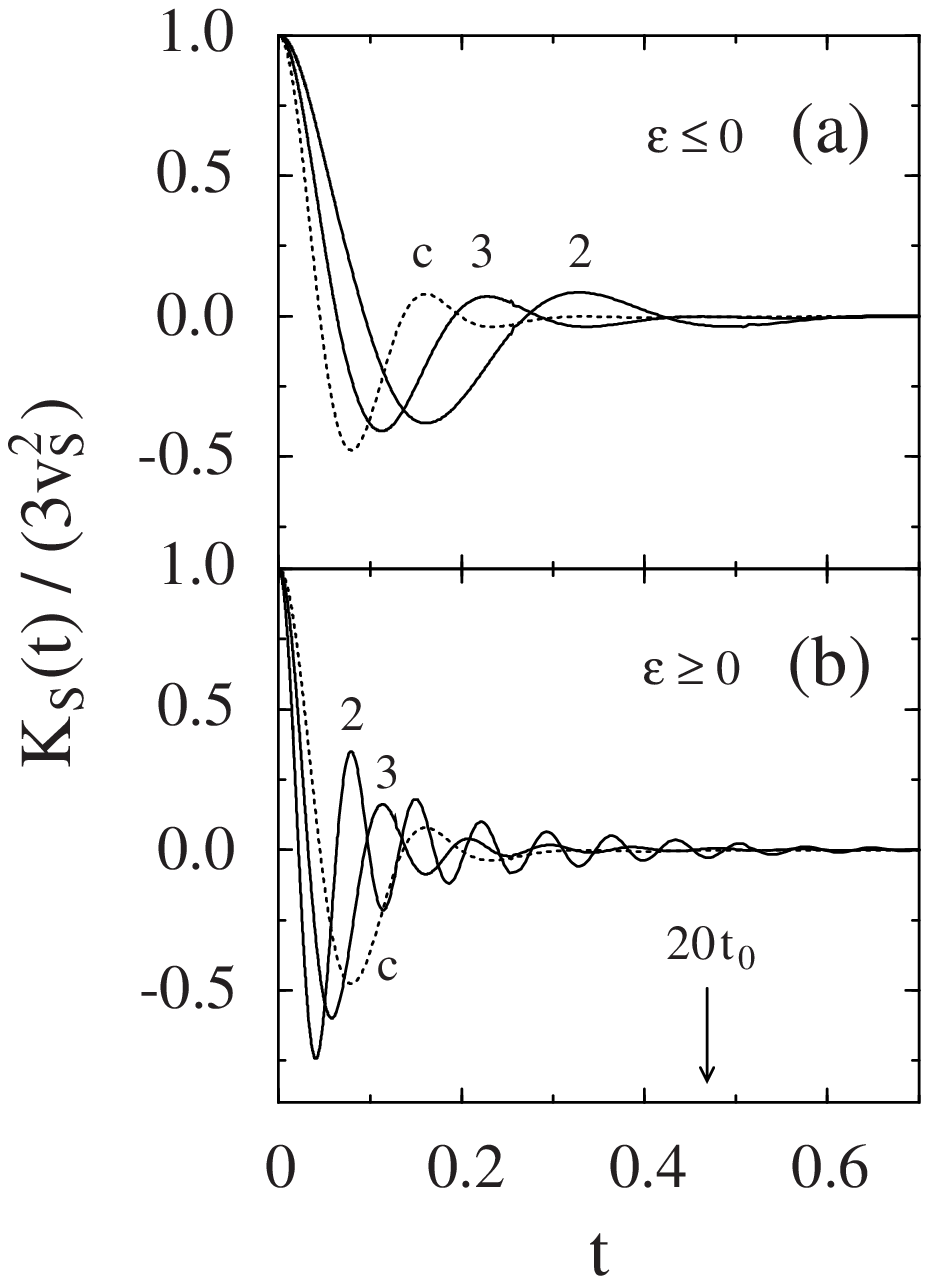}}}
\caption{Normalized velocity-correlation functions 
$K_S(t)/(3 v_S^2)$ for a tagged particle of the hard-sphere system 
(HSS). 
The dotted lines with label c refer to the critical packing fraction
$\varphi_c \approx 0.516$. 
The full lines with labels $n=2$ and $3$ 
are calculated for
distance parameters 
$\epsilon = (\varphi-\varphi_{c})/\varphi_{c} = \mp 10^{- n/3}$ 
for the liquid  ($\epsilon < 0$)
and for the glass ($\epsilon > 0$), respectively.
Here and in some of the following figures an arrow
marks the time $20 t_0$, where $t_{0} = t_{0}^{\rm H} = 0.0236 (d/v)$ is the time
scale from Eqs.~(\protect\ref{eq:phi-A-critical}) and (\ref{eq:t0}) 
for the critical decay. The units of
length and of time are chosen here and in the following figures so
that the particle diameter $d$ and the thermal velocities $v =
v_S$ are unity. }
\label{fig:VACF}
\end{figure}

If a tagged particle would experience a mere Newtonian friction
force, the velocity
correlations would decay exponentially, 
$K_S (t) \propto \exp[- (t / \tau)]$. 
The cage effect in dense liquids manifests itself by a qualitatively
different behavior, namely by oscillatory
variations with a decay of $K_S (t)$ to negative 
values~\cite{Hansen86}. 
Figure~\ref{fig:VACF}(a) demonstrates this phenomenon.
With increasing density, the crossover time to
negative values shortens and the damping of the oscillations
increases.
A Green-Kubo formula relates
the particle diffusivity $D_S$ to the zero-frequency velocity
spectrum: $D_S = (1/3) \int_0^\infty dt \, K_S (t)$~\cite{Hansen86}. 
Negative contributions to $K_S (t)$ 
reduce the diffusivity with increasing $\varphi$. 
From Eq.~(\ref{eq:GLE-K}), one gets
$D_S$ as inverse of the zero-frequency
spectrum of the relaxation kernel~\cite{Goetze91b}:
$D_{S} = 1 / {\textstyle \int_{0}^{\infty}} dt \, m_{S}(t)$. 
From Eqs.~(\ref{eq:GLE-Delta-S}), 
one obtains for the long-time
asymptote of the MSD:
$\lim_{t \to \infty} \Delta_S (t) / t = D_S$~\cite{Hansen86}.

For glass states, density fluctuations arrest for long times:
$\phi_q^S (t \to \infty) = f_q^S > 0$. The Lamb-M\"ossbauer
factor $f_q^S$ is to be evaluated from the mode-coupling
functional ${\cal F}_q^S$ in 
Eqs.~(\ref{eq:MCT-x}) via the equation
$f_q^S / (1 - f_q^S) = {\cal F}_q^S [f_k^S, f_p]$~\cite{Bengtzelius84}. 
It approaches unity for $q$ tending to zero.
A localization length $r_S$ can be introduced to characterize the
width of the $f_q^S$--versus--$q$ curve: 
$f_{q}^{S} = 1 - (q r_S)^2 + O(q^4)$. 
From Eq.~(\ref{eq:phi-x-qnull-a}),
one gets for the MSD: $\lim_{t \to \infty} \Delta_S (t) = r_S^2$. 
Using Eqs.~(\ref{eq:GLE-Delta-S}), 
one can express $r_S^2$ as inverse of the long-time limit of
the relaxation kernel~\cite{Goetze91b}:
$r_{S}^{2} = 1 / m_{S}(t \to \infty)= 
1 / {\cal F}^{S}[f_{k}^{S}, f_{p}]$. 
If the density increases, the localization length $r_S$ decreases.
As a result, the frequency of the oscillations of the particles in
their frozen cages increases, as is shown in 
Fig.~\ref{fig:VACF}(b). 
But, in contrast to what is found for liquid states, the damping
of the oscillations decreases upon compression. This 
reflects the formation of anomalous oscillation
peaks in the density-fluctuation spectra, which have properties
of the so-called boson peaks of liquids and glasses~\cite{Goetze00}.

\begin{figure}
\centerline{\scalebox{0.60}{\includegraphics{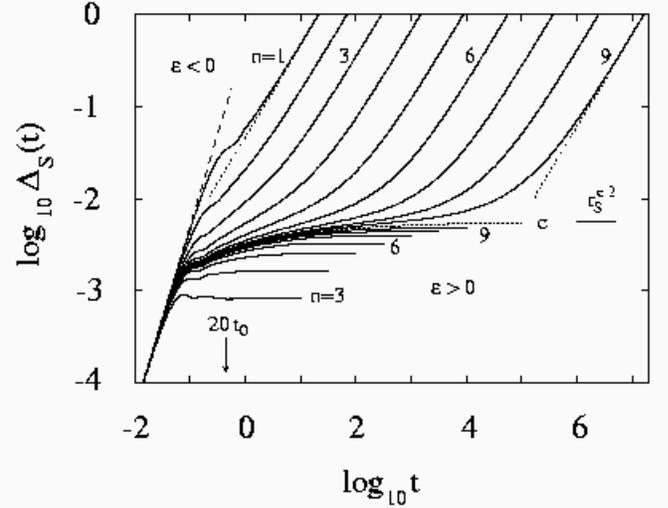}}}
\caption{Double logarithmic presentation of 
$\Delta_S(t) = \delta r_S^2(t) /6$ 
for the mean-squared displacement (MSD) $\delta r_S^2 (t)$
for a tagged particle of the HSS. The dotted line with label c
refers to the critical packing fraction $\varphi_c$ and the full
ones to $\epsilon = \pm 10^{- n/3 }$. 
The straight dashed line with slope 2 exhibits the
ballistic asymptote $(v_S t)^2 /2$. The straight dotted lines with
slope 1 exhibit the long-time asymptotes $D_S \, t$ of the two
liquid curves for $n = 1$ and 9. The horizontal line
marks the square of the localization length at $\varphi = \varphi_c$: 
$r_S^{c \, 2} = 0.00557 \, d^{2}$. }
\label{fig:MSD}
\end{figure}

The ideal liquid-glass transition implies a transition from a
regime with particle diffusion for $\varphi < \varphi_c$ to one
with particle localization for $\varphi \geq \varphi_c$. The
former is characterized by $D_S > 0$ and $1 / r_S = 0$ and the
latter by $D_S = 0$ and $1 / r_S > 0$. The subtleties of the
glass-transition dynamics occur outside
the transient regime. 
They can be discussed best on logarithmic scales as in 
Fig.~\ref{fig:MSD}. 
For very short times, say $t \leq t_0$, interaction effects are unimportant
and $\lim_{t \to 0} \Delta_S (t) /t^2 = v_S^2 / 2$
reflects ballistic motion. For times larger than $t_0$, the
cage effect leads to a suppression of $\Delta_S (t)$ below the
short-time asymptote. For such large times that $\delta r_S^2
(t) / d^2$ reaches unity, the MSD approaches the diffusion
asymptote, $\Delta_S (t) \approx D_S \, t$, as is shown by the dotted
straight lines drawn 
for the curves with labels $n = 1$
and $n = 9$. Upon increasing $\varphi$ towards $\varphi_c$, the
diffusivity decreases towards zero. 
Figure~\ref{fig:D-rS-scaling} shows that there
holds the power law 
$D_S^{1/\gamma} \propto \, \mid \epsilon \mid$ for
$\mid \epsilon \mid \, < 0.1$.

The lowest line in Fig.~\ref{fig:MSD} deals with the same glass state 
$\varphi = 1.1 \varphi_c$, which was considered in 
Fig.~\ref{fig:VACF}(b) for the label $n = 3$. 
For this density, there is no obvious glassy dynamics.
Rather, $\Delta_S (t)$ has approached its long-time limit $r_S^2$
after the oscillations have disappeared for $t \approx 1$.
Decreasing $\varphi$ towards $\varphi_c$, the softening of the
glass manifests itself by an increase of the localization length
$r_S$. At the transition point $\varphi = \varphi_c$, the critical
value $r_S^c = 0.0746 \, d$ is reached. 
This upper limit for $r_{S}$ is consistent with Lindemann's melting
criterion~\cite{Bengtzelius84}. 
Using Eq.~(\ref{eq:f-A}) for $f_{q}$ and $f^{S}_{q}$ and 
substituting these formulas into ${\cal F}^{S}[f^{S}_{k},f_{p}]$,
it follows that the glass
instability at $\varphi_c$ causes a $\sqrt{\sigma}$--anomaly
for the localization length,
\begin{equation}
r_{S}^{2} = r_{S}^{c \, 2} - h_{S} \, \sqrt{\sigma} \, / 
\sqrt{1-\lambda} + O(\sigma),
\label{eq:rS-2}
\end{equation}
where $r_S^{c \, 2} = 0.00557 \, d^{2}$ and 
$h_S = 0.0116 \, d^{2}$. 
Figure~\ref{fig:D-rS-scaling} 
demonstrates that the leading asymptotic formula
accounts for the $r_S^2$--versus--$\varphi$ dependence for
$\epsilon \leq 0.01$. But the data for $n \leq 4$, i.e., for
$\epsilon \geq 0.05$, are not anymore described by the
$\sqrt{\sigma}$--law. The range of applicability for the
asymptotic description of $r_S^2$ is remarkably smaller than the
one for the corresponding description of $f_q$ for intermediate
wave numbers~\cite{Franosch97}.

The glass curve for $\epsilon = 0.01$, shown in Fig.~\ref{fig:MSD} with label
$n = 6$, exhibits a decay between the end of the transient
oscillations and the arrest at $r_S^2$ which is stretched over a
time interval of about two orders of magnitude. A similar
two-decade interval is needed for the liquid curve with label $n =
6$ to reach the critical value $r_S^{c \, 2}$.
After crossing $r_{S}^{c \, 2}$, two further decades of an upward bent
$\log_{10} \Delta_S (t)$--versus--$\log_{10} t$ variation is
exhibited before the diffusion asymptote is reached. 
The indicated slow and stretched time variation is referred to as glassy
dynamics. 

\subsection{The structural-relaxation regime}
\label{subsec:relaxation-regime}

For times outside the transient regime, 
say $t \ge C^{*}t_{0}$, the density correlators
can be written in the form $\phi_q (t) = \phi_q^* (t / t_0)$. 
Here $t_0$ is the scale introduced in 
Eq.~(\ref{eq:phi-A-critical}). 
The functions $\phi_q^*$
are determined uniquely by the mode-coupling functional 
${\cal F}_q$, i.e., they are given by the equilibrium structure. 
This holds for all choices of the regular kernels in 
Eq.~(\ref{eq:GLE-v}), in particular for the two models defined by
Eqs.~(\ref{eq:R-model-a}) and 
(\ref{eq:R-model-b})~\cite{Chong01,Franosch98,Goetze92,Fuchs99b}.
Corresponding results
hold for the density correlators of the tagged 
atom~\cite{Goetze92} and of the molecule~\cite{Chong01}. 
The solutions of the specified MCT model for colloids are
completely monotone~\cite{Goetze95b}. A function $F (t)$, defined
for $t > 0$, is called completely monotone if all derivatives
exist and
\begin{mathletters}
\label{eq:dist-1}
\begin{equation}
(-\partial/\partial t)^{n} F(t) \ge 0, \quad
n = \mbox{0, 1, } \cdots.
\label{eq:dist-1-a}
\end{equation}
According to Bernstein's theorem~\cite{Gripenberg90}, this property
is equivalent to the existence of a distribution $\rho (\gamma)
\geq 0$ so that
\begin{equation}
F(t) = \int_{0}^{\infty} e^{-\gamma t} \, \rho(\gamma) \, d\gamma. 
\label{eq:dist-1-b}
\end{equation}
Thus, one can write for $t \geq C^* t_0$:
$\phi_{q}(t) = \int_{0}^{\infty} e^{- \gamma (t/t_{0})} \,
\rho_{q}(\gamma) \, d\gamma$ with
$\rho_{q}(\gamma) \ge 0$,
i.e., the functions $\phi_{q}^{*}$ deal with relaxation. 
\end{mathletters}
Corresponding formulas hold for the solute correlators
$\phi^{x}_{q}(t)$.
Representative examples for $\rho_q (\gamma)$ are discussed in
Ref.~\cite{Franosch99b}. 

From Eqs.~(\ref{eq:phi-x-qnull}),
one gets
$C_{1}(t) = C_{1}^{*}(t/t_{0})$ for $t \ge C^{*} t_{0}$,
where $C_{1}^{*}$ is completely monotone, and also 
\begin{equation}
\Delta_{x}(t) = \Delta_{x}^{*}(t/t_{0}), \quad
t \ge C^{*}t_{0}, \quad
x = S, C, A.
\label{eq:Delta-x-star}
\end{equation}
Here, the functions $C_{1}^{*}$ and $\Delta_{x}^{*}$ are determined by the 
equilibrium structure.
With Eq.~(\ref{eq:Delta-S-prop}) one 
can express the velocity correlator in
terms of the structure function 
$F (\tau) = - \partial_\tau^2 \Delta_S^* (\tau)$:
\begin{equation}
K_{S}(t) = - 3 \, F(t/t_{0}) \, / \, t_{0}^{2}, \quad
t \ge C^{*} t_{0}.
\label{eq:KS-star}
\end{equation}
It can be shown that the function $F$ is completely
monotone.
The proof does not provide further insight and 
is delegated to Appendix~\ref{appen:A}.

In Ref.~\cite{Franosch98}, the curves for $\phi_q (t)$
have been compared with the ones for 
$\phi_q^* (t / t_0)$. 
The interval for structural relaxation was larger for the
colloid model than that for the model with the underlying
Hamiltonian dynamics. 
Thus, the structural relaxation interval can be
identified as the one, where the specified curves for the two
models collapse. Figure~\ref{fig:MSD-N-B} 
exhibits such comparison for the MSD.
For $\mid \epsilon \mid \, \leq 0.01$, 
the curves agree within the accuracy of the drawing.
This holds, provided $t \geq 20 t_0$, i.e., $C^* \approx 20$. The
result is nearly valid also for distance parameters as large as
$\mid \epsilon \mid \, = 0.1$. But there is a small offset between
the full and the dashed liquid curves for $x = 1$. This
means, that there is a smooth drift of $t_0$ with
changes of $\varphi$, which is different for the
hard-sphere colloid and for the conventional HSS. 

In Fig.~\ref{fig:VACF-N-B}, 
the rescaled velocity correlator $t_0^2 \, K_S (t) / 3$
for the critical packing fraction is shown as full line. This
diagram is an extension and magnification of the dotted lines from
Fig.~\ref{fig:VACF} for $t \geq 0.2$. 
The dashed line is the analogous result
$t_0^{{\rm B} \, 2} \, \partial_t^2 \Delta_S^{\rm col}(t)$ calculated for
the colloid model. 
The latter function is 
$(t_0^{\rm B} D^{S}_0)^2 F^{\rm col}(t)$, 
where the completely monotone function $F^{\rm col}(t)$
was introduced in Appendix~\ref{appen:A} in connection with 
Eqs.~(\ref{eq:A8-a}) and (\ref{eq:A9}). 
According to Eq.~(\ref{eq:KS-star}), the two curves should collapse on the
function $F (t /t_0)$. This is the case for $t > 20 t_0$ within
small error margins. 
The curves demonstrate stretched relaxation to zero, which cannot be
adequately represented on linear scales. It is shown also that
oscillatory motion tends to mask glassy relaxation.
From now on, the discussion will focus on the 
structural-relaxation regime $t \ge 20 t_{0}$.

\subsection{Scaling-law descriptions}

In this section, it shall be examined how well
the leading-order asymptotic results from Sec.~\ref{subsec:universal}
can account quantitatively for the MSD. 
Let us start with $\Delta_S (t)$ for $\varphi =
\varphi_c$. This result for the critical dynamics, i.e. the dotted
line in Fig.~\ref{fig:MSD}, is reproduced as full line in the semi-logarithmic
presentation in Fig.~\ref{fig:MSD-critical}. 
The transient dynamics for $t \leq 20 t_0$
accounts for about 45\% of the total increase of $\Delta_S (t)$
from zero to the long-time asymptote $r_S^{c \, 2}$. The
structural relaxation needed to approach $r_S^{c \, 2}$ up to 5\%,
i.e. 50\% of the total increase, is stretched over a large interval
of about four orders of magnitude time variation. The
leading-order formula for the MSD at the transition point 
is in analogy to Eq.~(\ref{eq:phi-A-critical}):
\begin{equation}
\Delta_{S}(t) = r_{S}^{c \, 2} - h_{S} \, (t_{0}/t)^{a}, \quad
\varphi = \varphi_{c}, \quad t \gg t_{0}.
\label{eq:Delta-S-critical}
\end{equation}
The dashed line demonstrates that Eq.~(\ref{eq:Delta-S-critical})
describes --
within a 5\% error margin -- about 25\% of the total increase of
$\Delta_S (t)$. There remains the large part of the
structural-relaxation interval between 
$20 t_0$ and $t^* = 792 t_0$ 
which is not adequately accounted for. About
half of the structural-relaxation increase of $\Delta_S (t)$ is
outside the range of applicability of the leading-order
asymptotic formula. 
Extending Eq.~(\ref{eq:Delta-S-critical}) by
inclusion of the leading correction, one gets a
description of the MSD up to errors of order $(t_0 / t)^{3a}$:
$\Delta_S (t) = r_S^{c \, 2}- h_S (t_0 /t)^a + k_S (t_0 / t)^{2a}$. 
From Ref.~\cite{Fuchs98} one deduces $k_S = 0.0143 \, d^{2}$. 
The dashed-dotted line shows how inclusion
of the correction term expands the range of the analytic
description. 

The first scaling law for the MCT-bifurcation dynamics,
Eqs.~(\ref{eq:phi-A-first}), implies with
Eq.~(\ref{eq:phi-x-qnull-a})
\bea
& &
\Delta_{S}(t) = r_{S}^{c \, 2} - h_{S} \, \sqrt{\mid \sigma \mid} \,
g_{\pm}(t/t_{\sigma}),
\nonumber \\
& & \quad \quad
\sigma \, \gtrless\,  0, \quad \mid \sigma \mid \, \ll 1, \quad 
t \gg t_{0}.
\label{eq:Delta-S-scaling}
\eea
For the liquid states with $\sigma < 0$, it describes how the 
$\Delta_S (t)$--versus--$t$ curve crosses the plateau $r_S^{c \, 2}$. For
the glass states with $\sigma > 0$, it describes the approach
towards the arrest at $r_S^2 = \Delta_S (t \to \infty)$. 
The control-parameter independent functions 
$g_{\pm} (\hat t)$ for the HSS value of $\lambda$
are discussed in Fig.~10 of Ref.~\cite{Franosch97}. 
The dashed lines in Fig.~\ref{fig:MSD-beta} exhibit 
Eq.~(\ref{eq:Delta-S-scaling}) for three liquid states. 
They agree with the MCT solutions 
within a 5\% error margin within the intervals
marked by diamonds. 
Within these intervals, $\Delta_S (t)$ increases
from about $0.0043 \, d^{2}$ to about $0.0086 \, d^{2}$. 
Formulas like 
Eqs.~(\ref{eq:phi-A-critical}) and (\ref{eq:Delta-S-critical}) 
are the basis for the
derivation of the MCT-scaling laws. Therefore, it follows from
Fig.~\ref{fig:MSD-critical} 
that the part of the structural
relaxation regime between $20 t_0$ and $t^*$ remains outside the
range of validity of Eq.~(\ref{eq:Delta-S-scaling}).

For small rescaled times $\hat t = t/t_\sigma$, 
Eq.~(\ref{eq:Delta-S-scaling}) reproduces 
Eq.~(\ref{eq:Delta-S-critical}) for $\sigma \to 0$.
The r.h.s.~of Eq.~(\ref{eq:Delta-S-scaling}) 
becomes independent of $\sigma$ and agrees
with Eq.~(\ref{eq:Delta-S-critical}). 
This explains, why the dashed
lines in Fig.~\ref{fig:MSD-beta} 
for $x = 3$ and 4 collapse for $t \leq 10$ and
why the corresponding diamonds are located near $t^*$. 
For $x = 2$, the separation parameter $\sigma$ is already so large,
that $g_{\pm} (t^* / t_\sigma)$ differs remarkably from $(t_\sigma
/ t^*)^a$. Therefore, $\Delta_S (t)$ does not reach the $t^{-a}$
asymptote for $t \approx t^*$, and the
corresponding diamond shifts away from $t^*$. The $x = 3$ curve
follows the critical asymptote from 
Eq.~(\ref{eq:Delta-S-critical}) for a time interval of less
than two decades. 
Such interval would not be large
enough for a compelling experimental confirmation of the $t^{-a}$ law. 
To identify the $t^{-a}$ law in its pure form
for the model under study, 
$\mid \epsilon \mid$ must not exceed $10^{-4}$.

\begin{figure}
\centerline{\scalebox{0.60}{\includegraphics{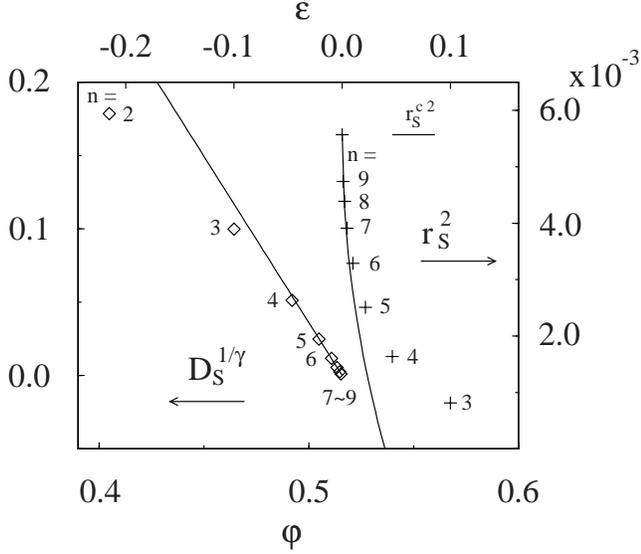}}}
\caption{The diamonds are the values for $D_S^{1 / \gamma}$ with
the HSS exponent $\gamma = 2.46$ for the tagged particle
diffusivities $D_S$ determined for the liquid curves in 
Fig.~\protect\ref{fig:MSD}.
The straight line is the function 
$\Gamma (\varphi_{c} - \varphi)$, $\varphi \leq \varphi_c$, 
with $\Gamma$ chosen so
that the line goes through the data point for $n = 9$. 
The crosses exhibit the
square of the localization length, $r_S^2$, determined for the
glass curves in Fig.~\protect\ref{fig:MSD}. 
The full line exhibits the leading
asymptotic law 
\protect$r_S^{c \, 2} - h_S \, \sqrt{\sigma} \, / \, \sqrt{1-\lambda}$, 
Eq.~(\protect\ref{eq:rS-2}). }
\label{fig:D-rS-scaling}
\end{figure}

\begin{figure}
\centerline{\scalebox{0.60}{\includegraphics{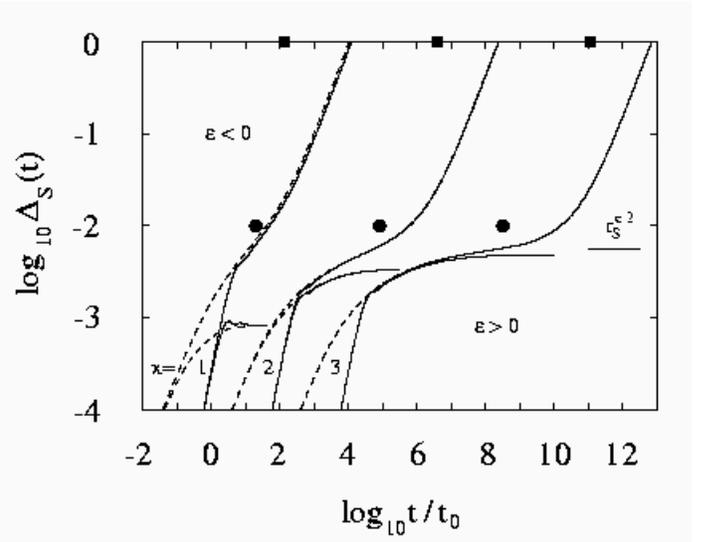}}}
\caption{Double logarithmic presentation of the MSD for a tagged
particle of the HSS as function of the reduced time $t / t_0$ for
distance parameters $\epsilon = \pm 10^{- x}$;
$x = 1$, 2, 3. The full lines reproduce the results from 
Fig.~\protect\ref{fig:MSD} with
labels $n = 3, 6, 9$ and $t_{0} = t_{0}^{\rm H}$ from 
Eq.~(\protect\ref{eq:t0}). 
The dashed lines are the corresponding results for the colloid 
model specified in
the text~\protect\cite{Fuchs98} with $t_{0} = t_{0}^{\rm B}$ from 
Eq.~(\protect\ref{eq:t0}). 
The curves for $x = 2$ and 3 are shifted by two and four decades,
respectively, to the right to avoid overcrowding. The full dots
and squares mark the time scales $t_\sigma$ and $t_\sigma^\prime$,
respectively, defined in 
Eqs.~(\ref{eq:phi-A-first-c}) and (\ref{eq:phi-A-von-b}). }
\label{fig:MSD-N-B}
\end{figure}

\begin{figure}
\centerline{\scalebox{0.65}{\includegraphics{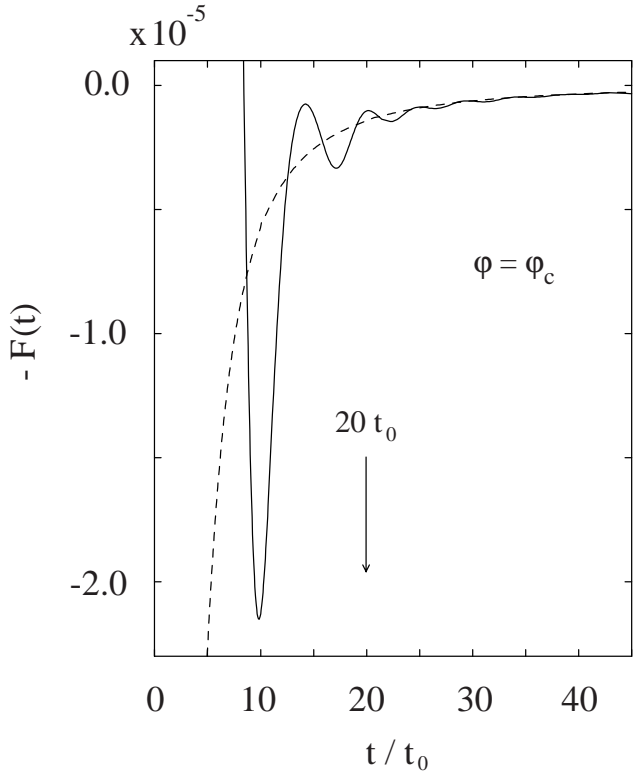}}}
\caption{Rescaled velocity correlators 
$- F(t) = t_0^2 \, K_S (t) /3 = t_0^2 \, \partial_t^2 \Delta_S (t)$ 
as function of the rescaled time
$t / t_0$ for a tagged particle of the HSS at the critical packing
fraction $\varphi = \varphi_c$. 
The full line refers to the model
for a Hamiltonian dynamics with $t_{0} = t_{0}^{\rm H}$
from Eq.~(\ref{eq:t0}) and the dashed
one to the colloid model defined in the text with 
$t_{0} = t_{0}^{\rm B}$ from Eq.~(\protect\ref{eq:t0}). 
The full line is a rescaling of the dotted curves
in Fig.~\protect\ref{fig:VACF}, where
values for $t \leq 0.2$, i.e. $t/t_{0} < 8.5$, 
are not reproduced.} 
\label{fig:VACF-N-B}
\end{figure}

\begin{figure}
\centerline{\scalebox{0.65}{\includegraphics{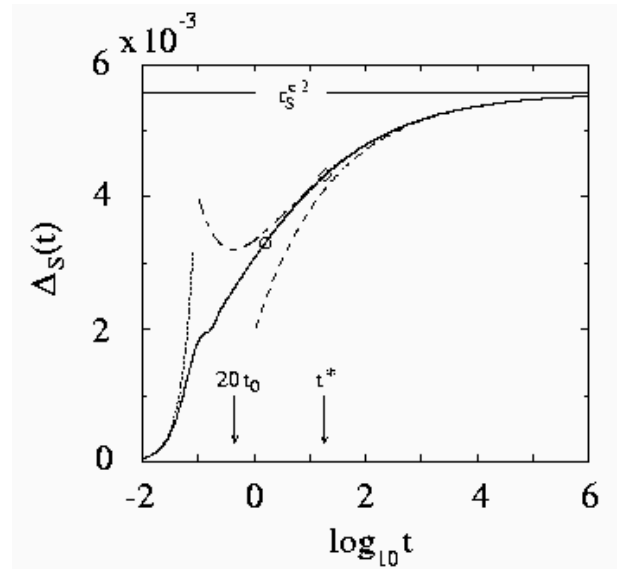}}}
\caption{$\Delta_S (t)$ for a tagged particle of the
HSS at the transition point $\varphi = \varphi_c$ (full line), the
leading-asymptotic expansion, Eq.~(\protect\ref{eq:Delta-S-critical}) 
(dashed line), and the
leading-plus-next-to-leading-asymptotic expansion $\Delta_S
(t) = r_S^{c \, 2} - h_S (t_0 / t)^a + k_S (t_0 / t)^{2a}$
(dashed-dotted line). The diamond and circle mark the times 
$t^* = 18.7 = 792 t_{0}$ and 
$t^{**} = 1.55 = 65.7 t_{0}$ 
where the full line differs by 5\%
from the dashed and dashed-dotted line, respectively. The
horizontal line marks the long-time asymptote $r_S^{c \, 2}$. The
dotted line exhibits the ballistic asymptote 
$\frac{1}{2} (v_S t)^2$.} 
\label{fig:MSD-critical}
\end{figure}

\begin{figure}
\centerline{\scalebox{0.55}{\includegraphics{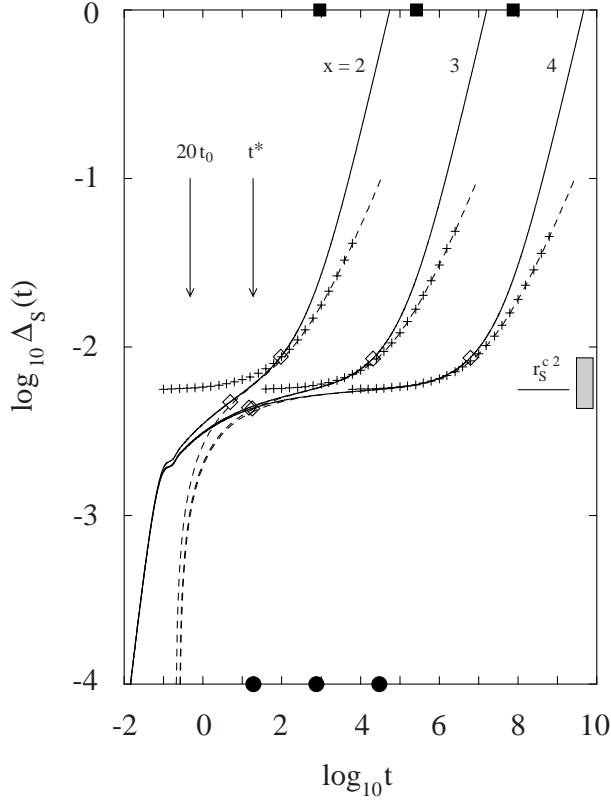}}}
\caption{The full lines show $\Delta_S (t)$ 
for packing fractions $\varphi$ given by 
$\epsilon = (\varphi - \varphi_c) /\varphi_c = - 10^{-x}$, 
$x = 2$, 3, 4. 
The dashed lines are the first-scaling-law descriptions 
by Eq.~(\protect\ref{eq:Delta-S-scaling}). 
The diamonds mark the points where the dashed lines differ from
the full ones by 5\%. Within these intervals, $\Delta_S (t)$ varies
between about 0.0043 and 0.0086 as is indicated by the shaded bar.
The crosses exhibit the von Schweidler law,
Eq.~(\protect\ref{eq:Delta-S-von}).
The filled circles and squares mark the times
$t_\sigma$ and $t_\sigma^\prime$, respectively,
defined in Eqs.~(\protect\ref{eq:phi-A-first-c}) and 
(\protect\ref{eq:phi-A-von-b}).}
\label{fig:MSD-beta}
\end{figure}

For large rescaled times $\hat t$, the master function for the
glass approaches 
$g_{+}(\hat{t} \to \infty) = 1 / \sqrt{1 - \lambda}$,
and Eq.~(\ref{eq:Delta-S-scaling}) 
reproduces Eq.~(\ref{eq:rS-2}). 
According to the preceding paragraphs, this explains the solutions 
for the glass provided 
the long-time limit $r_S^2$ is located in the shaded interval of
Fig.~\ref{fig:MSD-beta};
and this is demonstrated in Fig.~\ref{fig:D-rS-scaling}.

Function $g_{-}(\hat{t})$ for the liquid is zero for $\hat t_- =
0.704$. Thus, $\Delta_S (\hat t_- t_\sigma) = r_S^{c \, 2}$ and the
interval for the increase of $\Delta_{S}(t)$ to the plateau
value $r_S^{c \, 2}$ expands proportional to $t_\sigma$ if
$\varphi$ increases towards $\varphi_c$. 
For large $\hat{t}$, 
one gets von-Schweidler's law
\begin{equation}
\Delta_{S}(t) = r_{S}^{c \, 2} + h_{S} \, (t/t_{\sigma}^{\prime})^{b},
\quad \sigma \to -0, \quad t_{\sigma} \ll t \ll t_{\sigma}^{\prime}. 
\label{eq:Delta-S-von}
\end{equation}
Therefore, the long-time end of the range of applicability of 
Eq.~(\ref{eq:Delta-S-scaling}) 
expands proportional to $t_\sigma^\prime$, as is indicated by
the filled squares in Fig.~\ref{fig:MSD-beta}. 
Formula (\ref{eq:Delta-S-von}) is exhibited by the
crosses. These approach the plateau $r_S^{c \, 2}$ for
$t \ll t_\sigma$ and the dashed scaling-law lines for $t \gg
t_\sigma$. Since $t_\sigma^\prime / t_\sigma \to \infty$ for
$\mid \epsilon \mid \, \to 0$, the time interval for the von-Schweidler-law
description expands with decreasing 
$\mid \epsilon \mid$.

\begin{figure}
\centerline{\scalebox{0.60}{\includegraphics{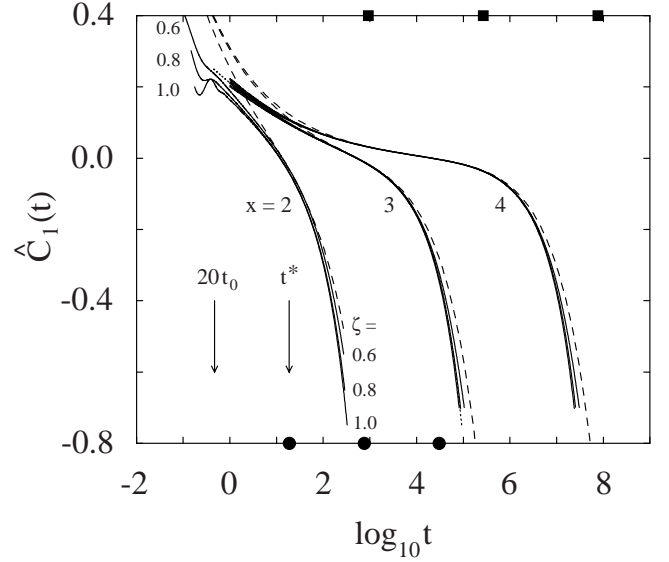}}}
\caption{The full lines are dipole correlators $C_1 (t)$ rescaled
to $\hat C_1 (t) = [C_1 (t) - f_1^c ] / h_1$ for the elongation
parameters $\zeta$ = 0.6, 0.8 and 1.0 (from top to bottom)
where
$f_{1}^{c} = 0.769$, $0.905$, $0.955$ and
$h_{1}     = 0.46$,  $0.19$,  $0.09$, respectively.
The distance parameters are $\epsilon = - 10^{-x}$, $x = 2$, 3, 4 
(compare text). 
The filled circles and squares mark the times $t_\sigma$
and $t_\sigma^\prime$, respectively, for the three 
packing fractions.
The dashed lines exhibit the first-scaling-law asymptotes
$\sqrt{\mid \sigma \mid} g_- (t / t_\sigma)$. The dotted line is
the MSD for a tagged particle for $x = 3$ rescaled to $\hat
\Delta_S (t) = [r_S^{c \, 2} - \Delta_S (t)] / h_S$.}
\label{fig:C1-beta}
\end{figure}

Equation~(\ref{eq:phi-A-first-a}) formulates
the factorization theorem for 
$\delta \phi_A (t) = \phi_A (t) - f_A^c$: 
in a leading-order expansion for small $\delta\phi_A$, the deviation
$\delta \phi_{A}(t)$ 
of the correlator from the plateau value $f_A^c$ factorizes in a
control-parameter-independent amplitude $h_A$ and a
function $G (t)$.
The function $G (t)$
is the same for all variables $A$ and describes the 
time-and-control-parameter dependence of $\phi_A (t)$
by a scaling law, Eq.~(\ref{eq:phi-A-first-b}).
This theorem can be tested by
identifying the time interval and the range of distance
parameters $\epsilon$ for which the diagrams for $\hat \phi_A (t) =
\delta\phi_A (t) / h_A$ collapse with $G(t)$. 
Figure~\ref{fig:C1-beta} demonstrates such a test for 
the dipole correlators $C_1 (t)$ 
for three values of the elongation parameter $\zeta$.
The markers for
$20 t_0, \, t^*, \, t_\sigma$ and $t_\sigma^\prime$ have been
added to ease a comparison with Fig.~\ref{fig:MSD-beta}. 
Obviously, the
scenario for the plateau crossing is the same for $C_1 (t)$ as
discussed above for $\Delta_S (t)$. To corroborate this
conclusion, the rescaled result for the MSD, 
$\hat \Delta_S (t) = [r_S^{c \, 2} - \Delta_S (t)] / h_S$ 
for $\epsilon = - 0.001$, has been
added to the figure. The $x = 2$ results show that for negative
$\hat{C}_1(t)$ the full lines follow the sequence
$\zeta =$ 0.6, 0.8 and 1.0 from top to bottom, 
where the latter two curves are very close
to each other. The same behavior is observed for positive 
$\hat{C}_1(t)$. This observation exemplifies a general 
implication of the leading corrections to the factorization 
theorem~\cite{Franosch97}.

A possibility for the definition of a
characteristic time scale $\tau_A$ for the $\alpha$-process of
variable $A$ is given by the time needed to complete 95\% of the
total decay from the plateau $f_{A}^{c}$ to the equilibrium
value zero, i.e., $\phi_A (\tau_A) = f_A^c / 20$. 
Within the range
of validity of the second scaling law, 
Eq.~(\ref{eq:phi-A-second}), the $A$-dependent scales
are coupled in the sense that 
$\tau_A = \tilde \tau_A \, t_\sigma^\prime$. 
Here $\tilde \tau_A$ is an $A$-specific
control-parameter-independent factor determined by
$\tilde \phi_A (\tilde \tau_A) = f_A^c / 20$. 
Applying these results to
the dipole correlator, one gets
$C_{1}(t) = \tilde{C}_{1}(\tilde{t})$
for $|\sigma| \ll 1$ and
$t_{\sigma} \ll t$
where $\tilde{t} = t/t_{\sigma}^{\prime}$.
For the $\alpha$-scale factors one finds 
$\tilde \tau_1$ = 18.8 (8.73, 2.66) for $\zeta$ = 1.0 (0.8, 0.6).
The description of the $\alpha$-process for elongation
parameter $\zeta$ = 0.8 is demonstrated in 
Fig.~8 of Ref.~\cite{Goetze00c}. 

Using the second scaling law for the tagged-particle-density
correlator, 
one gets from Eq.~(\ref{eq:phi-x-qnull-a})
the second scaling law for the $\alpha$-process of the MSD:
\begin{equation}
\Delta_{S}(t) = \tilde{\Delta}_{S}(\tilde{t}), \quad
\tilde{t} = t/t_{\sigma}^{\prime}, \quad 
\mid \sigma \mid \, \ll 1, \quad t_{\sigma} \ll t.
\label{eq:Delta-S-second}
\end{equation}
An $\alpha$-relaxation time $\tau_{S}$ shall be defined by that time,
where the diffusion asymptote $D_S \, t$ is reached within
5\%, i.e., $\Delta_S (\tau_S) = 1.05 \, D_S \, \tau_S$. 
One gets 
$D_S = \tilde D_S /t_\sigma^\prime$ and 
$\tau_S = \tilde \tau_S \, t_\sigma^\prime$, 
where $\tilde{D}_{S}$ and $\tilde{\tau}_{S}$ are to be determined from
$\lim_{\tilde{t} \to \infty} \tilde \Delta_S (\tilde t) / \tilde t =
\tilde{D}_{S}$ and
$\tilde \Delta_S (\tilde \tau_S) / \tilde \tau_S = 1.05 \tilde D_S$. 
One finds
$\tilde D_S = 0.0171$ and $\tilde \tau_S = 11.6$. 
One gets in particular 
$D_S \propto 1/t_\sigma^\prime \propto (\varphi_c - \varphi)^\gamma$, and
Fig.~\ref{fig:D-rS-scaling} 
demonstrates how the diffusivity
approaches this power-law asymptote for $\varphi$ increasing to
$\varphi_c$. 
The asymptotic description of $\Delta_S(t)$ by 
Eq.~(\ref{eq:Delta-S-second})
is demonstrated in Fig.~9 of Ref.~\cite{Fuchs98}.

For times of order $t_\sigma$, the relative corrections to the
first scaling law are of order $\sqrt{\mid \sigma \mid}$. 
For times of order $t_\sigma^\prime$, the corrections to the 
second scaling law are of order $\mid \sigma \mid$. Therefore, the 
second scaling law holds for larger separations $\mid \sigma \mid$ than
the first one~\cite{Franosch97}. For example, even for the large
value $\epsilon = - 0.1$, $D_S^{1/\gamma}$ differs from the linear
asymptote by only 15\%, as is demonstrated by the $n = 3$ result
in Fig.~\ref{fig:D-rS-scaling}. 
The corrections to Eq.~(\ref{eq:Delta-S-second}) increase if $t$ decreases
towards $t_\sigma$. But for $t \approx t_{\sigma}$,
the description by the first scaling
law becomes valid, which provides the leading corrections to 
Eq.~(\ref{eq:Delta-S-second}). 
The descriptions by the two
scaling laws overlap. Together, they provide a complete
description of the dynamics for $t \geq t^*$. This holds provided
$\mid \epsilon \mid$ is small enough, as is demonstrated in 
Fig.~\ref{fig:MSD-vs-zeta}
for $\epsilon = - 10^{-3}$ for 
$\Delta_S (t)$ and $\Delta_C (t)$ for
$\zeta =$ 0.8, and for $\Delta_A (t)$ for 4 values of $\zeta$.
From the analogous figure for $\epsilon = - 10^{-2}$ one concludes
that the scaling-law description accounts for the MSD
quantitatively for $t \geq t^*$ and 
$\mid \epsilon \mid \, \leq 0.01$. 
For larger distance parameters, corrections to scaling
become visible.

\subsection{Rotation-translation-coupling effects}

It might be adequate to start the discussion of 
rotation-translation-coupling effects by two side remarks. 
Firstly, it was shown that the correlators for
the dipole and the quadrupole dynamics for $\zeta = 0.8$ are in
semi-quantitative agreement with the experimental data for
propylene carbonate~\cite{Goetze00c}. Thus, the results 
to be discussed for $\zeta \geq 0.6$ 
can be considered as relevant for the interpretation of
glass-forming van-der-Waals liquids.
Secondly, the system
under study exhibits two glass phases for $\varphi \ge \varphi_{c}$.
There is a critical elongation $\zeta_{c}$ so that 
for $\zeta > \zeta_c$
correlations of the molecule's axis arrest for long times as do all
other correlations of variables characterizing the structure. In
particular $C_1 (t \to \infty) = f_1 > 0$. But for $\zeta \leq
\zeta_c$, dipole correlations exhibit ergodic behavior, i.e. $C_1
(t \to \infty) = 0$. Precursor effects of this glass-glass
transition at $\zeta_c$ disturb the
standard transition scenario~\cite{Chong01,Goetze00c}. This is the
reason why $\Delta_A (t)$ for $\zeta = 0.4$
is remarkably different from $\Delta_{A}(t)$ for
the other $\zeta$ shown in Fig.~\ref{fig:MSD-vs-zeta};
for $\zeta = 0.4$, the elongation is too close to $\zeta_c = 0.380$. 
For example, for $\zeta = 0.4$,
the corrections to von Schweidler's law, which shift the dashed
lines on to the full ones for $t_\sigma < t < t_\sigma^\prime$,
are negative, while they are positive for the other cases.

The glass-transition of the HSS is driven by density
fluctuations with wave numbers near the first-peak position of
the structure factor $S_q$ close to $q = 7.0 / d$.
For the scale factor for the $\alpha$-process of these
fluctuations, defined by 
$\tilde \phi_q (\tilde \tau_q) = f_q^c / 20$, 
one gets $\tilde{\tau}_{7.0} = 6.0$.
For the $\alpha$-relaxation of the
tagged-particle-density correlations for the same wave vector, one
gets a similar number $\tilde \tau_{7.0}^S = 5.0$. 
One concludes that the $\alpha$-scale factor
$\tilde{\tau}_{S} = \tau_{S}/t_{\sigma}^{\prime} = 11.6$ 
for the approach of $\Delta_S (t)$ to the
diffusion limit is in the range within which relevant density
fluctuations decay to zero. 
The corresponding interval for the crossover from the end of
the von-Schweidler decay to the beginning of diffusion is
2.2 decades. This is shown by the lowest curve in 
Fig.~\ref{fig:MSD-vs-zeta} and
should be considered as the normal behavior for the
density-fluctuation dynamics in simple systems. The MSD for the
molecule's center behaves quite similarly;
$\Delta_{C} (t)$ for $\zeta = 0.8$ is nearly indistinguishable
from $\Delta_S (t)$.
However, Fig.~\ref{fig:MSD-vs-zeta} 
demonstrates also that $\Delta_{A}(t)$ behaves differently.
For $\zeta$ = 1.0 (0.8, 0.6), the above specified crossover
intervals are 3.3 (3.1, 2.8) decades wide. This means, the
crossover intervals of $\Delta_{A}(t)$ are larger than those of 
$\Delta_S (t)$ or $\Delta_C (t)$ by factors of about 13 (7, 4),
and so are the $\alpha$-scales $\tau_{A}$ compared to
$\tau_{S}$ or $\tau_{C}$. The reason
is the reorientational contribution to the MSD of the constituent
atom, i.e., 
the second term on the r.h.s.~of Eq.~(\ref{eq:Delta-A-decom}). 
There are two effects.
First, for 
$t/t_{\sigma}^{\prime} \gg \tilde{\tau}_{1}$, 
the dipole correlations are decayed to
zero, i.e., the molecule's axis is distributed on the unit sphere
with a constant probability density.
Therefore, there is the positive offset $X = (\zeta d)^{2}/12$
between the two functions referring to the atom and the center.
$\Delta_{A}(t) - \Delta_{C}(t) = X$ for 
$\tilde{t} = t/t_{\sigma}^{\prime} \ge \tilde{\tau}_{1}$
as is demonstrated in the linear
representation of the $\alpha$-process master functions
in Fig.~\ref{fig:MSD-vs-zeta-2}.
The slow decay of the relative offset
$X / D_{C}t$ explains that the ratio 
$\tilde{\tau}_{A} / \tilde{\tau}_{C}$
is larger than unity and this the more the larger $\zeta$ is.
Second, steric hindrance for reorientations increases with
$\zeta$.
For $\zeta > 0.8$, the $\alpha$-scale $\tilde{\tau}_{1}$ 
for dipole relaxations becomes comparable to or larger than 
$\tilde{\tau}_{C}$.
Therefore, there appears a interval
$1 < t/t_{\sigma}^{\prime} < \tilde{\tau}_{1}$,
after the end of the von-Schweidler-law regime and before the
beginning of the diffusion regime, 
where the $\Delta_{A}(t)$--versus--$t$ diagram,
in contrast to the $\Delta_{C}(t)$--versus--$t$ one,
exhibits a curvature.
This is shown for $\zeta=0.8$ and 1.0 in 
Fig.~\ref{fig:MSD-vs-zeta-2}.

If a hard sphere gets expanded to a dumbbell with a small
elongation, the molecule's center gets restricted more tightly in
its cage. Therefore, provided $\zeta$ is small, 
the localization length $r_C^c$ is smaller
than $r_S^c$ and it decreases with increasing $\zeta$. 
For $\zeta \approx 1.0$, the localization of one constituent atom
of the molecule
restricts the motion of its partner. Therefore, $r_C^c$ is smaller
than $r_S^c$ also for large elongations. Counter intuitively, the
theory does not lead to a monotonic interpolation between the
specified limits. The $r_C^{c \, 2}$--versus--$\zeta$ diagram in 
Fig.~\ref{fig:D-rA-vs-zeta} 
exhibits an oscillation and $r_C^c$ exceeds $r_S^c$ by about 10\% for
$\zeta$ near 0.6. 

For $\zeta \leq \zeta_c$, one gets $C_1 (t \to \infty) = 0$. If
one neglects the variation of the 
$r_C^{c \, 2}$--versus--$\zeta$ curve,
Eq.~(\ref{eq:Delta-A-decom}) 
leads to $r_A^{c \, 2} = r_S^{c \, 2} + (\zeta d)^2 / 12$.
The dotted line in Fig.~\ref{fig:D-rA-vs-zeta} 
shows that this formula explains the
increase of $r_{A}^{c}$
for $\zeta$ increasing up to $\zeta_c$. 
For $\zeta$ increasing above
$\zeta_c$, the decrease of $1 - f_1$ and of $r_C^c$ explains the
decrease of $r_A^c$.

It seems plausible that the number of paths for the molecule
through the system's configuration space decreases with increasing
elongation. Therefore, one might expect that the diffusivity 
$D_A = D_C$ is smaller than $D_S$ and decreases with increasing
$\zeta$. 
Figure~\ref{fig:D-rA-vs-zeta} confirms this expectation for small and for
large elongations. But, strangely, the calculated diffusivity 
is not a monotone function of $\zeta$; and for 
$\zeta \approx 0.6$,
$D_{C}$ exceeds $D_S$ by about 30\%.

\section{CONCLUSIONS}
\label{sec:4}

MCT predicts that the long-time parts of the 
density-correlation functions 
$\phi_{q}(t)$ of the solvent and $\phi_{q}^{x}(t)$ of the solute
deal with relaxation in the sense that they
are superpositions of Debye-relaxation processes
$\exp[ - \gamma(t/t_{0})]$.
They deal with
structural dynamics in the sense that the weight functions
$\rho_{q}(\gamma)$ and $\rho^{x}_{q}(\gamma)$ 
for these superpositions are uniquely determined by the
equilibrium structure. 
The corresponding dynamics is therefore referred to as structural relaxation.
Within the structural-relaxation regime, the subtleties of the normal-liquid
dynamics merely enter via the value for an over-all time scale $t_{0}$,
which is defined with the aid of the critical decay law, 
Eq.~(\ref{eq:phi-A-critical}).
The structural-relaxation interval was estimated for the 
hard-sphere system (HSS) under study as $t > 20 t_{0}$ 
(Sec.~\ref{subsec:relaxation-regime}, Fig.~\ref{fig:MSD-N-B}).
The specified properties of the MCT dynamics are asymptotic ones,
valid for large times near the transition point.
The estimation of the range of validity depends on the accuracy
level required and also on the particular function under discussion
Differentiation with respect to time, as is considered in 
Eq.~(\ref{eq:Delta-S-prop}) for the velocity correlator $K_{S}(t)$,
enhances oscillation features, i.e. deviations from relaxation behavior.
Figure~\ref{fig:VACF-N-B} demonstrates indeed, that one would estimate
the structural relaxation interval for $K_{S}(t)$ as $t > 40 t_{0}$,
i.e. somewhat more restrictive than for the MSD.

For a normal liquid, the velocity correlator is positive for
short times.
It is also positive for times exceeding a cross-over time $t_{h}$,
where it exhibits a hydrodynamic long-time tail proportional to
$t^{-3/2}$~\cite{Hansen86}.
But, upon approaching the glass-transition point, the regime for 
hydrodynamics shrinks to lower frequencies.
Therefore, the time $t_{h}$ diverges if the packing fraction $\varphi$
approaches the critical value $\varphi_{c}$.
The cage effect causes $K_{S}(t)$ to be negative for $t > 20t_{0}$.
Even more, the correlation of the particle velocity at time $t$
with the opposite of its initial value, say
$\Psi(t) = \langle {\vec v}_{S}(t) \cdot [- {\vec v}_{S}(t=0)] \rangle$,
is proportional to a completely monotone function 
$F(t/t_{0})$, Eq.~(\ref{eq:KS-star}).
The function $\Psi$ represents a relaxation process in the sense of 
Eq.~(\ref{eq:dist-1-b}), where the weight function $\rho(\gamma)$
is given by the structural function.
These results for the velocity correlator for $t > 20 t_{0}$ express
concisely the essence of glassy dynamics,
namely the relaxation caused by the cage effect and
determined by the Boltzmann factors for the equilibrium structure.

The MSD increases with time according to the characteristic two-step
pattern of the MCT-transition scenario.
There is a stretched approach towards a plateau value $r_{S}^{c \, 2}$
followed by arrest at the square of the localization length
$r_{S}^{2} < r_{S}^{c \, 2}$ within the glass or by the stretched
start of the $\alpha$-process within the liquid, Fig.~\ref{fig:MSD}.
The sensitive density dependence of the MSD near the plateau is described
by the universal formulas for the first scaling law, as is 
demonstrated in Fig.~\ref{fig:MSD-beta}.
The same holds for the dipole correlator $C_{1}(t)$.
This is shown in Fig.~\ref{fig:C1-beta} by scaling $C_{1}(t)$ for 
three values of the molecule's elongation on the same functions
$\sqrt{|\sigma|} \, g_{-}(t/t_{\sigma})$.
However, there is a large time interval at the beginning of the
structural-relaxation regime, $20 t_{0} < t < t^{*} \approx 800 t_{0}$,
where the universal leading-order-asymptotic formulas do not describe
the MSD, as is shown in 
Figs.~\ref{fig:MSD-critical} and \ref{fig:MSD-beta}.
The theory predicts similar results for all systems with a structure 
similar to that of the HSS, i.e. for all van-der-Waals liquids.
Obviously, it would be worthwhile to test by experiment or
molecular-dynamics simulation whether the specified prediction is
correct, in particular, whether MCT can reproduce properly the 
structural relaxation of the MSD outside the scaling-law regime.

\begin{figure}
\centerline{\scalebox{0.65}{\includegraphics{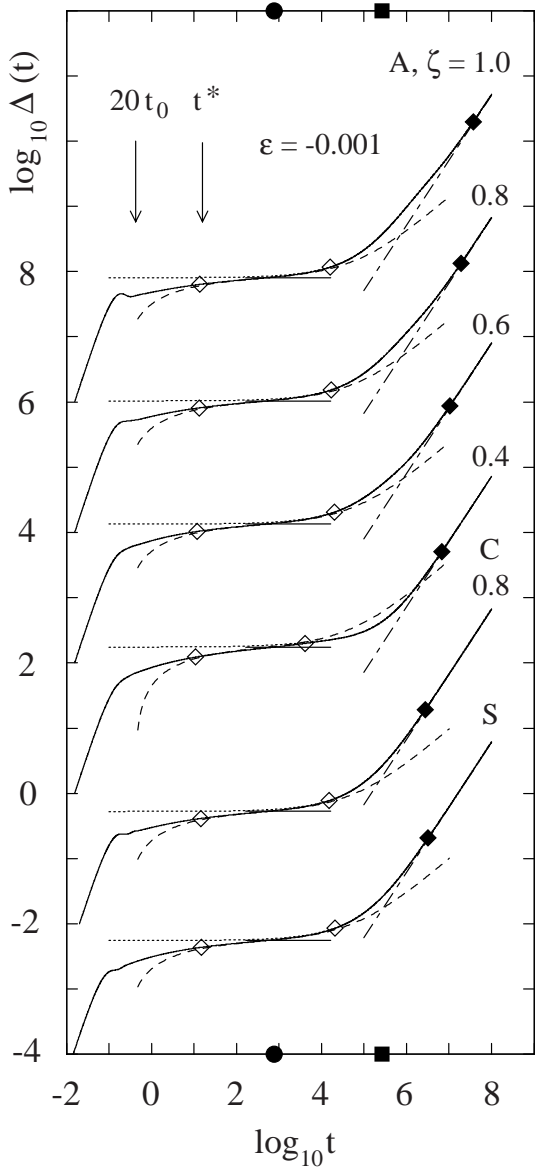}}}
\caption{$\Delta_S (t)$ and $\Delta_C (t)$ for $\zeta = 0.8$, and
$\Delta_A (t)$ for $\zeta$ = 0.4, 0.6, 0.8, 1.0 (full lines, from
bottom to top). Successive curves are shifted upwards by two
decades to avoid overcrowding. The distance parameter is $\epsilon
= - 10^{-3}$ and the corresponding times $t_\sigma$ and
$t_\sigma^\prime$ are marked by filled
circles and squares, respectively. The dashed lines are the
first-scaling-law asymptotes, 
Eq.~(\ref{eq:Delta-S-scaling}). 
The open diamonds mark the points where the dashed lines differ
from the full ones by 5\%. The straight dashed-dotted lines
exhibit the diffusion asymptotes $D_S t, \, D_C t$ and $D_A t$,
and the filled diamonds mark the position where
these differ from the full lines by 5\%. The dotted lines, which
coincide with the full ones for $t \geq 10^4$, exhibit the
second-scaling-law asymptotes, Eq.~(\ref{eq:Delta-S-second}).}
\label{fig:MSD-vs-zeta}
\end{figure}

\begin{figure}
\centerline{\scalebox{0.80}{\includegraphics{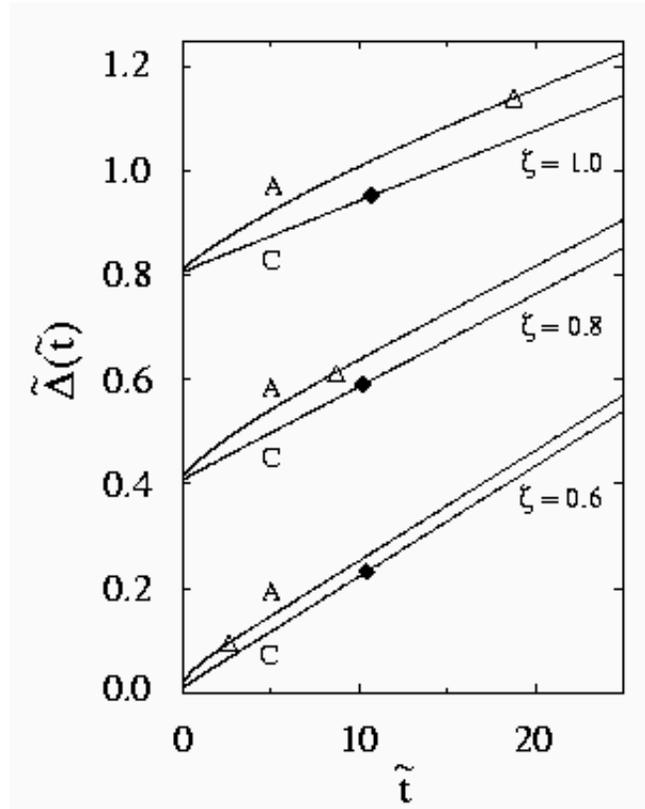}}}
\caption{
$\alpha$-relaxation master functions 
$\tilde{\Delta}$ as function of the rescaled
time $\tilde{t} = t / t_{\sigma}^{\prime}$ for
$\zeta = 1.0$, $0.8$ and $0.6$ (from top to bottom).
The curves for $\zeta = 0.8$ ($1.0$) are shifted upward
by 0.4 (0.8) in order to avoid overcrowding.
Curves with labels $C$ and $A$ refer to the MSD for the
molecule's center and for the constituent atom, respectively.
The filled diamonds mark the time $\tilde{\tau}_{C}$, where the
diffusion asymptote is reached within 5\%:
$\tilde{\Delta}_{C}(\tilde{\tau}_{C}) = 
1.05 \tilde{D}_{C} \tilde{\tau}_{C}$.
The corresponding time $\tilde{\tau}_{A}$ for the constituent atom
for $\zeta = 1.0$ (0.8, 0.6) is 
$\tilde{\tau}_{A} = 140$ (72, 39).
The triangles mark the time $\tilde{\tau}_{1}$, where
the dipole correlator has completed 95\% of its 
$\alpha$-decay: 
$\tilde{C}_{1}(\tilde{\tau}_{1}) / f_{1} = 0.05$.}
\label{fig:MSD-vs-zeta-2}
\end{figure}

The beginning of the $\alpha$-process of the MSD, i.e. the increase of 
$\Delta_{x}(t)$ above the plateau $r_{x}^{c \, 2}$, is described by
von Schweidler's law, Eq.~(\ref{eq:Delta-S-von}).
It is exhibited in Fig.~\ref{fig:MSD-vs-zeta} for $t > t_{\sigma}$
by the dashed lines.
The $\alpha$-process terminates in the diffusion law for long times,
exhibited in Fig.~\ref{fig:MSD-vs-zeta} by the straight dashed-dotted lines.
The $\alpha$-process follows well the second scaling law,
Eq.~(\ref{eq:Delta-S-second}), which is presented by the dotted lines.
The cross-over interval from the end of the von-Schweidler-law 
description (indicated by the open diamonds) to the beginning of the
diffusion process (indicated by the filled diamonds) for the MSD of
an atom is about 2 decades wide, as shown by the lowest curve 
in Fig.~\ref{fig:MSD-vs-zeta}.
The same is true for the MSD of the molecule's center.
But, the cross-over interval for the MSD of the constituent atom of the
molecule is much larger due to the rotation-translation coupling.
The expansion is more than an order of magnitude for $\zeta = 1.0$.
It would be valuable for an assessment of the MCT
for molecular systems if this prediction could be tested by 
molecular-dynamics simulation.

Bennemann {\it et al.}~\cite{Bennemann99}
have determined by molecular-dynamics simulation
the MSD for the monomer and for the center of a polymer
for a model of a glassy polymer melt. They identified for their
data the $\alpha$-process in the sense of MCT and interpreted it
consistently with the universal asymptotic formulas. They observed
the expansion of the crossover interval for the monomer
MSD relative to that for the center similar to what is shown in
Fig.~\ref{fig:MSD-vs-zeta} 
for $\zeta = 1.0$, and they attribute this expansion to the
glassy dynamics of the Rouse modes. These modes for the polymer's
internal degrees of freedom are thus identified as 
part of the $\alpha$-process~\cite{Bennemann99}. 
Figures~~\ref{fig:MSD-vs-zeta} and \ref{fig:MSD-vs-zeta-2}
corroborate their conclusions; albeit for the simplest polymer model only,
namely a rigid hard-sphere dimer. 

\begin{figure}
\centerline{\scalebox{0.60}{\includegraphics{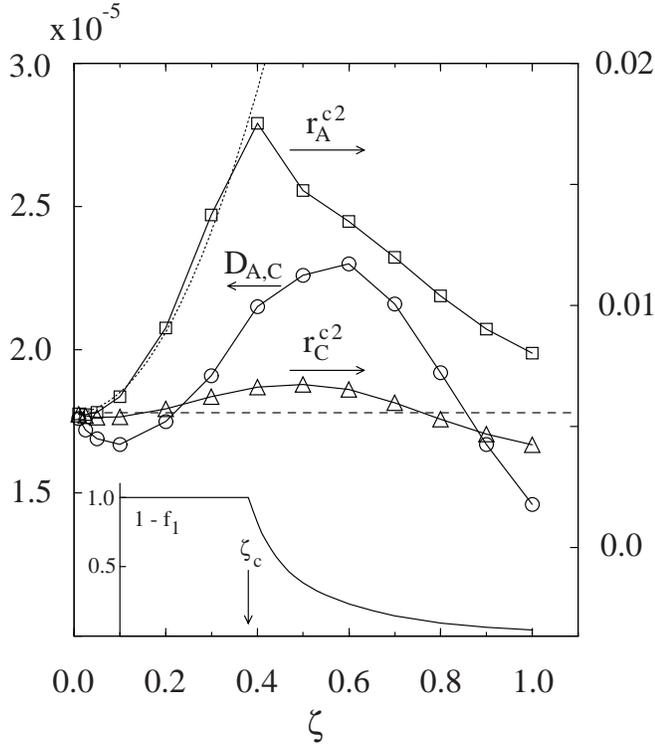}}}
\caption{Squares and triangles are the localization lengths
squared at the transition point $\varphi = \varphi_c$ as function
of the molecule's elongation $\zeta$ for 
the constituent atom ($r_{A}^{c \, 2}$) and
the center ($r_{C}^{c \, 2}$), respectively. 
The circles are the diffusivity 
$D_A = D_C$ of the molecule calculated for $\epsilon = - 0.01$. The
lines connecting the symbols are guides to the eye. 
The horizontal dashed line is a common one indicating 
both the values of the square of the localization length
$r_S^{c \, 2}$ (on the right scale)
and of the diffusivity $D_S$ (on the left scale) 
for a tagged particle of the HSS.
The dotted line is the function 
$r_S^{c \, 2} + (\zeta d)^2 / 12$
discussed in the text.
The inset exhibits $1 - f_1$ as function of $\zeta$, where 
$f_1 = C_1 (t \to \infty)$ is the long-time limit of
the dipole correlator $C_{1}(t)$.
The arrow indicates the critical elongation $\zeta_{c} = 0.380$
for a glass-glass transition.}
\label{fig:D-rA-vs-zeta}
\end{figure}

\begin{figure}
\centerline{\scalebox{0.65}{\includegraphics{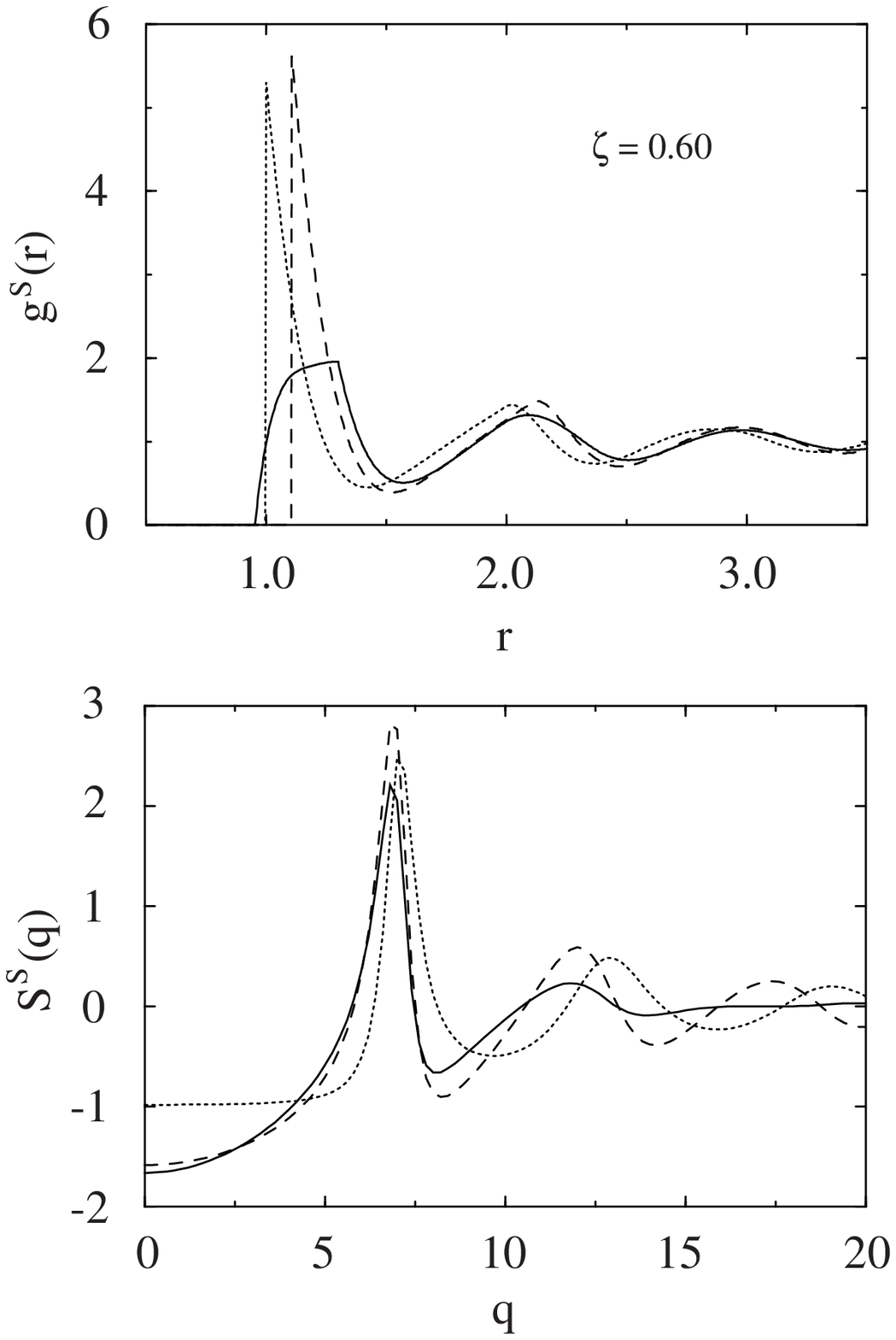}}}
\caption{
Pair distribution $g^{S}(r)$ as function of the distance $r$
between solvent particles and the center of the solute 
(upper panel) and the corresponding solute-solvent structure factor
$S^{S}(q)$ (lower panel) calculated within the 
Percus-Yevick theory for the 
critical packing fraction $\varphi_{c} = 0.516$.
The dotted lines refer to a tagged sphere of the same diameter,
$d_{S} = d$, as the one of the solvent particles.
The dashed lines refer to a spherical solute of diameter 
$d_{\rm eff} = 1.215 \, d$ (see the text).
The full lines exhibit the isotropic part of the distribution
$g_{0}^{S}(r)$ and the structure factor $S_{0}^{S}(q)$ 
for a dumbbell consisting of two fused hard spheres of
diameter $d$ with the elongation $\zeta = 0.6$~\protect\cite{Franosch97b}.}
\label{fig:gr-Sq}
\end{figure}

Let us consider a one-dimensional model, where
two hard spheres are restricted to move in a finite interval.
One calculates
easily the localization length $r_S$ of one of the spheres,
assuming the other sphere to move freely between the wall and its
partner. Similarly, one can calculate the localization length
$r_C$ of the center of a dumbbell built by the two spheres. 
One finds $r_C > r_S$, i.e., the freely
moving partner provides a stronger hindrance for the motion in the
cage than the bonded one. 
Possibly, the result $r_C^c > r_S^c$ shown in 
Fig.~\ref{fig:D-rA-vs-zeta} for
$\zeta \simeq 0.6$ is the analogue of this phenomenon for
one-dimensional localization.
To corroborate this reasoning, Fig.~\ref{fig:gr-Sq} exhibits
as full lines the isotropic part $g_{0}^{S}(r)$ of the solute-solvent
pair distribution function and the corresponding structure factor
$S_{0}^{S}(q)$ for $\varphi = \varphi_{c}$ for a dumbbell with 
$\zeta = 0.6$~\cite{Franosch97b}. 
The dashed lines exhibit the results for a sphere with diameter
$d_{\rm eff} = 1.215 \, d$ chosen such that its volume agrees
with that of the dumbbell. 
The well-known 
excluded-volume effects for this effective spherical
solute are larger than those for the tagged particle of the HSS.
In particular, the structure factor peak is higher.
Therefore, the localization length
$r_{\rm eff}^{c} = 0.0588 \, d$ for the effective spherical solute
is much shorter than the localization length
$r_{S}^{c} = 0.0746 \, d$ for the tagged particle of the HSS.
The ratio $r_{\rm eff}^{c} / r_{S}^{c}$ decreases with
increasing $\zeta$.
The pair distribution function of the effective sphere agrees 
reasonably well with $g_{0}^{S}(r)$ for the dumbbell for
$r > 1.5 \, d$.
Therefore, the two structure factors are close to each other
on the wings of the $S_{0}^{S}(q)$-peaks and also for small $q$.
However, packing cannot be done so efficiently around the
dumbbell than around a sphere of the same volume.
Therefore, the pair distribution function for the latter is 
much bigger for $r$ near $d_{\rm eff}$
than $g_{0}^{S}(r)$ for the dumbbell. 
This depletion effect leads to a reduction of
$\mid S_{0}^{S}(q) \mid$
at the peak position and also for larger wave numbers.
Hence, the depletion reduces the magnitude of the mode-coupling
coefficients.
As a result, the localization length 
$r_{0}^{c} = 0.0863 \, d$, calculated
by using MCT equations for a simple system
with only the isotropic part of the dumbbell structure factor,
is not only larger than 
$r_{\rm eff}^{c}$, but it is even larger than 
$r_{S}^{c}$.
The estimated $r_{0}^{c}$ is rather close to the true localization length
$r_{C}^{c} = 0.0808 \, d$ for the dumbbell's center.
Only if $\zeta$ becomes even larger,
mode-couplings to the non-isotropic parts of the density fluctuations
take over and compensate for the depletion effects.
This leads to the decrease of $r_{C}^{c}$ with increasing $\zeta$
for $\zeta > 0.6$, as shown in Fig.~\ref{fig:D-rA-vs-zeta}.

The above discussed depletion effect means, that the dumbbell is
surrounded by a liquid of lower averaged density than specified by 
the packing fraction.
Hence, the dumbbell moves in a complex, which effectively
has a larger distance from the glass-transition point than is specified
by $(\varphi_{c}-\varphi)/\varphi_{c}$.
Such lubrication phenomenon for the dumbbell in the complex 
might explain the increase
of the diffusivity exhibited in Fig.~\ref{fig:D-rA-vs-zeta}
for $\zeta \sim 0.6$.
Diffusion in a dense liquid is a collective phenomenon. A particle
can move over a distance comparable to its diameter only, if one
of its neighbors moves. This neighbor can move only if its
neighbor moves and so on. On the average, a flow pattern will be
built like one knows it for the motion of a sphere in an
incompressible ideal liquid. It was one of the original
motivations for the formulation of the MCT-equation of motion to
approximately treat the indicated back-flow phenomenon.
Molecular-dynamics simulation has given evidence that
a typical event underlying the mentioned averaged back-flow 
pattern might consist 
of quasi-one-dimensional motions of clusters of 
particles~\cite{Donati98}. It seems possible, that the bonding of two atoms
to a molecule stabilizes the cluster, provided the elongation
$\zeta d$ is smaller than the diameter of the ring formed by the
moving cluster. If this was the case, the diffusivity for the
molecule should be larger than that for a tagged atom. Hence, it
is not obvious that the non-monotone variation of the
diffusivity with changes of $\zeta$,
which is shown in Fig.~\ref{fig:D-rA-vs-zeta},
is a mere artifact of the approximations
underlying the presented theory. 

\acknowledgments

We thank M.~Fuchs, W.~Kob, M.~Sperl, 
W.~van Megen and Th.~Voigtmann for 
helpful and stimulating discussions.
S.-H.~C. acknowledges financial support from
JSPS Postdoctoral Fellowships for Research Abroad. 
This work was supported by 
Verbundprojekt BMBF 03-G05TUM.

\appendix

\section{MODE-COUPLING-THEORY MODELS FOR COLLOIDS}
\label{appen:A}

In this appendix, it will be shown how the MCT formulas for simple
systems can be specialized to models for the dynamics of colloids.
Within this frame, a representation of the velocity correlator in
terms of a completely monotone function $F^{\rm col}(t)$ 
shall be derived. To
begin, let us remember the definition of Laplace
transforms of functions of time, say $F (t)$, to functions of the
complex frequency $s$: 
$F (s) = \int_0^\infty dt \, \exp (- s t) \, F(t)$. 
Applying this transformation to $\phi_q (t)$, $M_{q}^{\rm reg}(t)$ 
and $m_q (t)$, one can rewrite
Eq.~(\ref{eq:GLE-v}) 
together with the initial conditions $\phi_q (t = 0) =
1$ and $\partial_t \phi_q (t = 0) = 0$ as a double-fraction
for $\phi_q (s)$:
\begin{equation}
\phi_{q}(s) = 1 \, / \,
\{  s + \Omega_{q}^{2} \, / \, [s + M_{q}^{\rm reg}(s) + 
\Omega_{q}^{2} \, m_{q}(s)] \}.
\label{eq:A1}
\end{equation}
The analogous result can be obtained for the tagged-particle
correlator
\begin{equation}
\phi_{q}^{S}(s) = 1 \, / \,
\{  s + (q v_{S})^{2} \, / \, [s + M_{q}^{S, {\rm reg}}(s) + 
(q v_{S})^{2} \, m_{q}^{S}(s)] \}.
\label{eq:A2}
\end{equation}
Equation~(\ref{eq:Delta-S-prop}) 
can be rewritten as a relation between the
Laplace transforms $\Delta_S (s)$ and $K_S (s)$ of $\Delta_S (t)$
and $K_S (t)$, respectively:
\begin{equation}
K_{S}(s) = 3s^{2} \, \Delta_{S}(s). 
\end{equation}
The small--$q$--expansion of Eq.~(\ref{eq:A2}) leads to
\begin{equation}
K_{S}(s) = 3 v_{S}^{2} \, / \, [s + M_{0}^{S, {\rm reg}}(s) +
v_{S}^{2} \, m_{S}(s)].
\label{eq:A4}
\end{equation}
Here $M_{0}^{S, {\rm reg}}(s) = \lim_{q \to 0} M_{q}^{S, {\rm reg}}(s)$,
and $m_S (s) = \lim_{q \to 0} q^2 m_q^S (s)$ 
is the transform of $m_S (t)$
from Eq.~(\ref{eq:MCT-x-qnull-a})~\cite{Goetze91b}.

In a colloid, there is a contribution to the fluctuating force due
to the interaction of the particles with the suspending liquid. 
This leads to contributions to the
rates $M_{q}^{\rm reg}(s)$ and $M_{q}^{S, {\rm reg}}(s)$ 
which are large compared to
the scale for the frequency $s$ one wants to consider. Therefore,
it is the conventional first approximation step of a theory for
colloid dynamics to coarse grain correlation functions
on time scales of the order of 
$1 / M_{q}^{\rm reg}(s)$ and $1 / M_{q}^{S, {\rm reg}}(s)$. 
For the correlators considered here, this means that 
$s + M_{q}^{\rm reg}(s)$ in Eq.~(\ref{eq:A1}), 
$s + M_{q}^{S, {\rm reg}}(s)$ in Eq.~(\ref{eq:A2}), 
and $s + M_{0}^{S, {\rm reg}}(s)$ in Eq.~(\ref{eq:A4})
are replaced by 
$\nu_q = M_{q}^{\rm reg}(s=0)$,
$\nu_q^S = M_{q}^{S, {\rm reg}}(s=0)$, and
$\nu_0^S = M_{0}^{S, {\rm reg}}(s=0) = \lim_{q \to 0} \nu_{q}^{S}$,
respectively. 
Let us indicate the various functions, obtained by this specialization
of MCT, by superscripts ``col''. The equations of motion, obtained by
the back transformation of the so modified 
Eqs.~(\ref{eq:A1}) and (\ref{eq:A2}),  
are~\cite{Franosch97,Fuchs98}:
\bea
& &
\tau_{q} \, \partial_{t} \phi_{q}^{\rm col}(t) + 
\phi_{q}^{\rm col}(t) 
\nonumber \\
& & \quad \quad \quad \quad
+ \,
\int_{0}^{t} dt^{\prime} \, m_{q}^{\rm col}(t-t^{\prime}) \,
\partial_{t^{\prime}} \phi_{q}^{\rm col}(t^{\prime}) = 0,
\label{eq:A5}
\\
& &
\tau_{q}^{S} \, \partial_{t} \phi_{q}^{S, {\rm col}}(t) + 
\phi_{q}^{S, {\rm col}}(t) 
\nonumber \\
& & \quad \quad \quad \quad
+ \,
\int_{0}^{t} dt^{\prime} \, m_{q}^{S, {\rm col}}(t-t^{\prime}) \,
\partial_{t^{\prime}} \phi_{q}^{S, {\rm col}}(t^{\prime}) = 0. 
\label{eq:A6}
\eea
Here $\tau_q = \nu_q / \Omega_q^2$ and $\tau_q^S = \nu_q^S / (q
v_S)^2$ are times characterizing conventional colloid dynamics.
The interaction potentials between the particles, which cause the
cage effect, are not altered by the introduction of the solvent.
Therefore, the expressions for the mode-coupling contributions to
the kernels, Eqs.~(\ref{eq:MCT-v}) and (\ref{eq:MCT-x}), 
keep their functional form 
\begin{equation}
m_{q}^{\rm col}(t) = {\cal F}_{q}[\phi_{k}^{\rm col}(t)], \quad
m_{q}^{S, {\rm col}}(t) = 
{\cal F}_{q}^{S}[\phi_{k}^{S, {\rm col}}(t), \phi_{p}^{\rm col}(t)].
\label{eq:A7}
\end{equation}
Equations (\ref{eq:A5})--(\ref{eq:A7}), together with the initial conditions
$\phi_q^{\rm col} (t = 0) = 1$ and $\phi_q^{S,{\rm col}} (t = 0) =
1$ define a unique solution with all the general properties of
correlation functions~\cite{Goetze95b}. Since the coarse
graining has altered the $s \to \infty$ asymptote of the correlator
transforms relative to that exhibited by 
Eqs.~(\ref{eq:A1}) and (\ref{eq:A2}), the
short-time behavior is altered, namely 
$\phi_q^{\rm col} (t \to 0) = 1 - |t| / \tau_q + \cdots$,
$\phi_q^{S,{\rm col}} (t \to 0) = 1 - |t| / \tau_q^S + \cdots$. 
Since $s$ was
dropped in the denominator of Eq.~(\ref{eq:A4}), the velocity-correlator
Laplace transform for the colloid dynamics does not tend to zero
for large $s$; rather $K_S^{\rm col} (s \to \infty) = 3 D^{S}_{0}$, with
the abbreviation $D^{S}_{0} = v_S^2 / \nu^{S}_0$. This suggests to rewrite
the coarse-grained velocity correlator as
\begin{mathletters}
\label{eq:A8}
\bea
& &
K_{S}^{\rm col}(s)/3 =
D^{S}_{0} - D_{0}^{S \, 2} \, F^{\rm col}(s),
\label{eq:A8-a}
\\
& &
F^{\rm col}(s) =
m_{S}^{\rm col}(s) \, / \, 
[1 + D^{S}_{0} \, m_{S}^{\rm col}(s)],
\label{eq:A8-b}
\\
& &
m_{S}^{\rm col}(t) =
{\cal F}^{S}[\phi_{k}^{S,{\rm col}}(t), \phi_{p}^{\rm col}(t)]. 
\label{eq:A8-c}
\eea

The solutions of Eqs.~(\ref{eq:A5})--(\ref{eq:A7}) 
are completely monotone~\cite{Goetze95b}. 
\end{mathletters}
Kernel $m_S^{\rm col} (t)$ is a combination with
positive coefficients of products of completely monotone functions
because of Eqs.~(\ref{eq:A8-c}) and (\ref{eq:MCT-x-qnull}), 
and therefore it is completely
monotone as well~\cite{Gripenberg90}. Because of Bernstein's
theorem, one can write $m_S^{\rm col} (s) = \int_0^\infty [s +
\gamma]^{-1} \rho (\gamma) d \gamma$ with a non-negative
distribution $\rho (\gamma)$. Therefore, there hold the following
four properties~\cite{Gripenberg90} for $m_S^{\rm col} (s)$: (i)
it is holomorphic for all complex $s$, except for negative real
numbers, (ii) $m_S^{\rm col} (s)^* = m_S^{\rm col} (s^*)$,
(iii) $m_S^{\rm col} (s) \to 0$ for $\mbox{Re} \,  s \to \infty$, and 
(iv) $\mbox{Im} \, m_S^{\rm col} (s) < 0$ for $\mbox{Im} \, s > 0$. 
One concludes furthermore
that $1 + D^{S}_0 m_S^{\rm col} (s) \not= 0$ for all $s$ except,
possibly, negative real values. From Eq.~(\ref{eq:A8-b}) 
one concludes that also $F^{\rm col} (s)$ exhibits the properties 
(i)--(iv). Hence,
$F^{\rm col} (s)$ is the Laplace transform of a completely
monotone function $F^{\rm col} (t)$ (Ref.~\cite{Gripenberg90},
chapter 5, theorem 2.6). The back transformation of 
Eq.~(\ref{eq:A8-a}) leads to the desired representation
\begin{equation}
K_{S}^{\rm col}(t)/3 = D^{S}_{0} \, \delta(t) - 
D_{0}^{S \, 2} \, F^{\rm col}(t),
\label{eq:A9}
\end{equation}
where $F^{\rm col} (t)$ obeys Eqs.~(\ref{eq:dist-1}). 

Let us add, that the back transformation of Eq.~(\ref{eq:A8-b}) leads to an
equation of motion determining $F^{\rm col} (t)$ from the kernel
$m_S^{\rm col} (t)$:
\begin{equation}
F^{\rm col}(t) = m_{S}^{\rm col}(t) - D^{S}_{0} \,
\int_{0}^{t} dt^{\prime} \,
m_{S}^{\rm col}(t-t^{\prime}) \, F^{\rm col}(t^{\prime}). 
\label{eq:A10}
\end{equation}
Integrating Eq.~(\ref{eq:A9}) twice with respect to the time and using 
Eq.~(\ref{eq:A10}), one gets an equation of motion for the MSD of the colloid
\begin{equation}
\Delta_{S}^{\rm col}(t) = D^{S}_{0} \,
\biggl[
t - \int_{0}^{t} dt^{\prime} \,
m_{S}^{\rm col}(t-t^{\prime}) \, \Delta_{S}^{\rm col}(t^{\prime})
\biggr]. 
\end{equation}
This result was derived originally along a different route~\cite{Fuchs98}. 
It implies $\lim_{t \to 0} \Delta_S^{\rm col} (t) / t = D^{S}_0$.

\newpage\noindent

\newpage\noindent

\newpage\noindent

\newpage\noindent

\newpage\noindent

\newpage\noindent

\newpage\noindent

\newpage\noindent

\newpage\noindent

\newpage\noindent

\newpage\noindent

\newpage\noindent

\begin{thebibliography}{10}

\bibitem{Hansen86}
J.-P. Hansen and I.~R. McDonald, {\em Theory of Simple Liquids}, 2nd ed.
  (Academic Press, London, 1986).

\bibitem{Megen98}
W. van Megen, T.~C. Mortensen, J. M{\"u}ller, and S.~R. Williams, Phys. Rev. E
  {\bf 58},  6073  (1998).

\bibitem{Kegel00}
W.~K. Kegel and A. van Blaaderen, Science {\bf 287},  290  (2000).

\bibitem{Weeks00}
E.~R. Weeks, J.~C. Crocker, A.~C. Levitt,
A. Schofield, and D.~A. Weitz,
Science {\bf 287},  627  (2000).

\bibitem{Kob95}
W. Kob and H.~C. Andersen, Phys. Rev. E {\bf 51},  4626  (1995).

\bibitem{Kaemmerer98}
S. K{\"a}mmerer, W. Kob, and R. Schilling, Phys. Rev. E {\bf 58},  2131
  (1998).

\bibitem{Kudchadkar95}
S.~R. Kudchadkar and J.~M. Wiest, J. Chem. Phys. {\bf 103},  8566  (1995).

\bibitem{Mossa00}
S. Mossa, R. {Di Leonardo}, G. Ruocco, and M. Sampoli, Phys. Rev. E {\bf 62},
  612  (2000).

\bibitem{Gallo96}
P. Gallo, F. Sciortino, P. Tartaglia, and S.-H. Chen, Phys. Rev. Lett. {\bf
  76},  2730  (1996).

\bibitem{Doliwa99}
B. Doliwa and A. Heuer, J. Phys.: Condensed Matter {\bf 11},  A277  (1999).

\bibitem{Horbach99}
J. Horbach and W. Kob, Phys. Rev. B {\bf 60},  3169  (1999).

\bibitem{Goetze91b}
W. G{\"o}tze,  in {\em Liquids, Freezing and Glass Transition}, edited by J.-P.
  Hansen, D. Levesque, and J. Zinn-Justin (North-Holland, Amsterdam, 1991), p.\
  287.

\bibitem{Schilling94}
R. Schilling,  in {\em Disorder Effects on Relaxational Processes}, edited by
  R. Richert and A. Blumen (Springer-Verlag, Berlin, 1994), p.\ 193.

\bibitem{Goetze99}
W. G{\"o}tze, J. Phys.: Condensed Matter {\bf 11},  A1  (1999).

\bibitem{Wuttke00}
J. Wuttke, M. Ohl, M. Goldammer, S. Roth, 
U. Schneider, P. Lunkenheimer, R. Kahn, B. Ruffl\'e,
R. Lechner, and M. A. Berg,
Phys. Rev. E {\bf 61},  2730  (2000).

\bibitem{Torre00}
R. Torre, P. Bartolini, M. Ricci, and R.~M. Pick, Europhys. Lett. {\bf 52},
  324  (2000).

\bibitem{Hinze00}
G. Hinze, D.~D. Brace, S.~D. Gottke, and M.~D. Fayer, J. Chem. Phys. {\bf 113},
   3723  (2000).

\bibitem{Gleim00}
T. Gleim and W. Kob, Eur.\ Phys.\ J.\ B {\bf 13},  83  (2000).

\bibitem{Sciortino01}
F. Sciortino and W. Kob, Phys. Rev. Lett. {\bf 86},  648  (2001).

\bibitem{Fuchs92b}
M. Fuchs, I. Hofacker, and A. Latz, Phys. Rev. A {\bf 45},  898  (1992).

\bibitem{Fuchs95}
M. Fuchs, Transp. Theory Stat. Phys. {\bf 24},  855  (1995).

\bibitem{Franosch97}
T. Franosch, M. Fuchs, W. G{\"o}tze, M.~R. Mayr, and A.~P. Singh,
Phys. Rev. E {\bf 55},  7153  (1997).

\bibitem{Fuchs98}
M. Fuchs, W. G{\"o}tze, and M.~R. Mayr, Phys. Rev. E {\bf 58},  3384  (1998).

\bibitem{Goetze00}
W. G{\"o}tze and M.~R. Mayr, Phys. Rev. E {\bf 61},  587  (2000).

\bibitem{Chong01}
S.-H. Chong, W. G{\"o}tze, and A.~P. Singh, Phys. Rev. E {\bf 63},  011206
  (2001).

\bibitem{Chandler72}
D. Chandler and H.~C. Andersen, J. Chem. Phys. {\bf 57},  1930  (1972).

\bibitem{Bengtzelius84}
U. Bengtzelius, W. G{\"o}tze, and A. Sj{\"o}lander, J. Phys. C {\bf 17},  5915
  (1984).

\bibitem{Franosch98}
T. Franosch, W. G{\"o}tze, M.~R. Mayr, and A.~P. Singh, J. Non-Cryst. Solids
  {\bf 235--237},  71  (1998).

\bibitem{Franosch97b}
T. Franosch and A.~P. Singh, J. Chem. Phys. {\bf 107},  5524  (1997).

\bibitem{Goetze92}
W. G{\"o}tze and L. Sj{\"o}gren, Rep. Prog. Phys. {\bf 55},  241  (1992).

\bibitem{Fuchs99b}
M. Fuchs and $\mbox{Th}$. Voigtmann, Philos.\ Mag.\ B {\bf 79},  1799  (1999).

\bibitem{Goetze95b}
W. G{\"o}tze and L. Sj{\"o}gren, J. Math. Analysis and Appl. {\bf 195},  230
  (1995).

\bibitem{Gripenberg90}
G. Gripenberg, S.~O. Londen, and O. Staffans, {\em Volterra Integral and
  Functional Equations}, Vol.~34 of {\em Encyclopedia of Mathematics and Its
  Applications} (Cambridge University Press, Cambridge, 1990).

\bibitem{Franosch99b}
T. Franosch and W. G\"otze, J.~Phys.~Chem.~B {\bf 103},  4011  (1999).

\bibitem{Goetze00c}
W. G{\"o}tze, A.~P. Singh, and $\mbox{Th}$. Voigtmann, Phys. Rev. E {\bf 61},
  6934  (2000).

\bibitem{Bennemann99}
C. Bennemann, J. Baschnagel, W. Paul, and K. Binder, Comp.~Theo.~Poly.~Sci {\bf
  9},  217  (1999).

\bibitem{Donati98}
C. Donati, J.~F. Douglas, W. Kob, S.~J. Plimpton, 
P.~H. Poole, and S.~C. Glotzer,
Phys. Rev. Lett. {\bf 80},  2338  (1998).

\end{thebibliography}
\end{document}